\documentclass[]{aa}
\usepackage{natbib}
\usepackage{epsfig,lscape,rotating}
\usepackage{color}
\usepackage{amsmath}
\usepackage{amssymb}
\usepackage{ulem}
\usepackage{hyperref}

\renewcommand{\vec}[1]{\mbox{\boldmath$#1$}}

\begin{document}

\title{3D radiative hydrodynamic simulations of protostellar collapse with H-C-O dynamical chemistry}

\author{Natalia Dzyurkevich \inst{1}, Beno\^it Commer\c{c}on \inst{2}, Pierre Lesaffre \inst{1}, Dimitry Semenov \inst{3} }
  \institute{Laboratoire de Radioastronomie Millim\'etrique, UMR 8112 du CNRS, \'Ecole Normale
  Sup\'erieure et Observatoire de Paris, 24 rue de Lhomond, 75231, Paris Cedex
  05, France
\and
\'Ecole Normale Sup\'erieure de Lyon, CRAL, UMR 5574 du CNRS, Universit\'e de Lyon I, 46 all\'ee d'Italie, 69364, Lyon Cedex 07, France
\and
 Max-Planck-Institut f\"ur Astronomie, K\"onigstuhl 17, 69117
Heidelberg, Germany
      }


\authorrunning{N. Dzyurkevich et al.}
\titlerunning{Time-dependent dynamical chemistry during the protostellar collapse}

\date{}

\abstract
{Combining the co-evolving chemistry, hydrodynamics and radiative transfer in protostellar collapse simulation is a major achievement, 
for both observations and theoretical developments in star formation studies. First, it allows a better link to observations, 
such as CO$_2$ and OH lines in the infrared. Second, a detailed knowledge of self-consistent chemical evolution for the main charge carriers (both gas species and dust grains) allows to correctly  estimate the rate and nature of magnetic dissipation in the collapsing core.}
{We aim to describe the chemo-dynamical evolution of collapsing dense cores using a reduced gas-grain chemical network, which describes mainly H-C-O chemistry. 
We study the effect of free-fall time and of dust properties on the chemical evolution, i.e. the dynamical versus static chemistry. }
{We present the results of 3D simulations of 1 M$_\odot$ isolated dense core collapse. The physical
setup includes radiative hydrodynamics and dynamical evolution of a chemical network. In order to perform those
simulations, we merged the multi-dimensional adaptive-mesh-refinement code \ttfamily{RAMSES}\rm~ and 
the thermo-chemistry Paris-Durham shock {\ttfamily (PDS)} code.
We simulate the formation of the first hydro-static core 
 and the co-evolution of 56 species, mainly describing H-C-O chemistry for
the sake of computational feasibility.
}
{
We have systematically tested the reduced chemical network against a well-establiched complex network. 
We show that by using a compact set of reactions,
one can get a pretty good match with a much more complex network. This saves computational time and enables the chemo-dynamical RHD simulations of the cloud collapse in 3D. 
Our main results are threefold. (a) We follow in detail the time-dependent formation of the first
hydro-static core, until the central temperature of about 800~K and density of about $10^{13}$~cm$^{-3}$
were reached. After that, we use the output physical structure and apply the static gas-grain chemistry.
with an extended set of reactions.
 We find that gas-grain chemistry post-processing can lead to one order
of magnitude lower CO gas-phase abundances compared to the reduced dynamical chemistry, with strongest effect during the
isothermal phase of collapse. 
(b)
 The duration of the isothermal phase (i.e. free-fall time) has little effect on the chemical abundances for our choice of the parameters.
For mean grain sizes of 1$\mu{}$m and larger, the gas-dust interaction timescales become longer than the representative
dynamical timescales, which affects the pace of molecular depletion and makes dynamical chemistry a must.
(c) Furthermore, dust mean size and size distribution have a strong effect on chemical
abundances and hence on the ionization degree and magnetic dissipation. We present the
ionisation profile for the collapsing cloud. }
{Dynamical chemical evolution is required to describe the CO gas phase abundance as well 
as the CO ice formation for the mean grain size larger then  1$\mu{}$m. 
The latter point is of primary importance regarding the recent observational and theoretical evidences 
of the dust grain growth in the envelope of the protostellar cores.
We conclude that proper accounting for
dust grain growth in the collapse simulations can be as important as coupling the collapse
with chemistry.}

\keywords{chemistry --- hydrodynamics --- radiative transfer}


\maketitle

\section{Introduction}

There is hardly any other astrophysical topic linked to so many research fields in 
astronomy as star formation which involves a large variety of physical processes. Star formation is a key process to study the evolution of galaxies, the physics of 
the interstellar medium, the structure of molecular clouds, and the birth and evolution of planetary systems. 
Among others, chemistry is expected to play a crucial role during the star formation process: it regulates the thermal budget of the gas via atomic and molecular emission lines, as well as the degree of ionization of the gas, including the grain charging. A precise description of the ionisation in star forming regions is required to follow accurately the coupling between the gas and magnetic fields, which has some important consequences regarding the long-standing problems of angular momentum and magnetic flux conservation in star formation.

Understanding the star formation process thus requires the coupling of
chemistry with radiative magnetohydrodynamics (MHD) models which combine the effects of gravity, hydrodynamics, radiative transfer, and magnetic fields. The aim of such new generation of numerical models integrating chemistry is threefold: follow the main molecular tracers of physical conditions (e.g. the lines of CO, C$^{18}$O, HCO$^+$), describe the coupling between gas and magnetic fields, and account for the thermal feedback of chemistry. In this study about protostellar collapse, 
we will focus on the first aspect, leaving the question of the coupling between gas and magnetic fields to subsequent paper.
%

Chemistry is first necessary to follow the evolution of the chemical tracers of physical conditions during the collapse, in order to compare models with 
observations and thus to verify or falsify the models.
 The chemistry 
of pre-stellar cores has been studied extensively \citep[e.g.][]{diFrancesco07,Bergin07}.
There have been several studies applying chemical modelling to either observed 
dense clouds or to more general models of isolated collapsing cores. 
Gas-grain chemical networks are usually massive and the 
resulting abundances  are obtained at high computational costs \citep{hin13}. 
For this reason many researchers choose to use analytical or semi-analytical 
models for the dynamical evolution 
\citep[e.g.][]{Aikawa08}.  
With this simplification, \citet{vis09,vis11} could follow
the chemo-dynamical evolution during protostellar collapse up to the formation of 
a disk around the protostar. Similarly, \citet{mar13} applied a large chemical network
to a static spherical dense cloud at temperature 10 K and at density of about  
$10^4$~cm$^{-3}$, adopting physical parameters close to the L1498 and L1517B prestellar cores. They explored the impact of dust size and of proper 
treatment of cosmic ray propagation on the chemistry, comparing their synthetic 
maps with available observations. The most advanced works, which combine the evolution of a full gas-grain chemical network together with 3D (radiation-)MHD models can be found in  
\citet{fur12} and \citet{hin13}. They calculated 
the chemical evolution of gas parcels linked to tracer particles, which described the different components of the collapsing core: envelope, pseudo-disk, disk, outflow, and the first hydrostatic core ~\citep[FHSC][]{lar69}. 

All these chemo-dynamical models are in fact 
post-processing of hydrodynamic models/simulations. It allows the best link to observations, reproducing accurately the chemical abundances for given physical parameters. But, the feedback of the chemical abundances onto radiative and magneto-hydrodynamical
evolution is neglected.
 Here, we are seeking a possibility to obtain both a reasonable quality of chosen chemical species and to have an opportunity to use the information about ionization and ion spatial distribution directly in the radiative and resistive MHD simulations.



Our ambition here is to implement a dynamical chemical evolution for the radiation-(M)HD models of 
protostellar collapse. We merge the \ttfamily{RAMSES }\rm code \citep{teys02} and the 
thermo-chemical Paris-Durham shock {\ttfamily PDS} code \citep{flower03,lesaffre05,lesaffre13,flower15}.
The long-term goal is twofold. First, we aim to reproduce well enough the CO abundance with a simple network, 
because CO lines are important for the gas cooling  during the collapse. Second, we aim to use the number densities of the  charge carries for dynamical calculations of the magnetic diffusivities in the collapse simulations in follow-up studies. Here we present the details on the R-MHD and chemical codes and the first parameter study. 
  
The paper is organised as follows. In section 2, we present the reduced chemical network we use for protostellar collapse and the coupling of chemistry and radiation-hydrodynamics in  \ttfamily{RAMSES}\rm. Section 3 presents an application to protostellar collapse in which we study the effect of dust grain properties and dynamical time on the chemical state within the collapsing cloud.  Section 4 concludes our work.


\section{Methods and codes: chemistry in \ttfamily{RAMSES}\rm}

In this section we present the chemo-dynamical solver we have integrated in the 
adaptive-mesh-refinement (AMR) code \ttfamily{RAMSES} \rm \citep{teys02}. 
We first describe the reduced chemical network we design for protostellar 
applications and confront it against a full gas-grain chemical model. 
Then we present our implementation of dynamical chemistry in  \ttfamily{RAMSES}\rm.

\subsection{The reduced chemical network\label{sec:chemmod}}

\subsubsection{Chemical species}

\begin{table}
\begin{tabular}{lll}
\hline
 Element & \ttfamily{ALCHEMIC}\rm~  &  \ttfamily{RAMSES}\rm   \\
\hline
\multicolumn{3}{c}{ Gas-phase elements}\\
$\rm H_2$  & 0.5                & 0.5    \\
$\rm He$   & 0.09               & 0.09  \\
$\rm C^+$  & $1.2\times10^{-4}$  & $1.2\times10^{-4}$  \\
$\rm O$    & $2.56\times10^{-4}$ & $2.56\times10^{-4}$   \\
$\rm N$    & $7.6\times10^{-5}$  & -  \\
$\rm S^+$  & $8\times10^{-8}$    & - \\
$\rm Si^+$ & $8\times10^{-9}$    & - \\
$\rm Na^+$ & $2\times10^{-9}$    & - \\
$\rm Mg^+$ & $7\times10^{-9}$    & - \\
$\rm Fe^+$ & $3\times10^{-9}$    & $3\times10^{-9}$ \\
$\rm P^+$  & $2\times10^{-10}$   & -  \\
$\rm Cl^+$ & $1\times10^{-9}$    &  - \\
\hline
\multicolumn{3}{c}{ Grain core elements}\\
$\rm O^{**}$  & - & $1.1\times10^{-4}$ \\
$\rm Si^{**}$ & - & $3.4\times10^{-5}$ \\
$\rm Mg^{**}$ & - & $3.7\times10^{-5}$ \\
$\rm Fe^{**}$ & - & $3.2\times10^{-5}$ \\
$\rm C^{**}$  & - & $7.3\times10^{-4}$ \\
\hline
$f_\mathrm{dg}$ & 0.01 & 0.01085 \\
\hline
\end{tabular}
\caption{ \label{tab:param2}  Elemental abundances used in \ttfamily{ALCHEMIC}\rm~ \citep{semenov10}  and
       in \ttfamily{RAMSES(PDS)}\rm. 
       Missing elements are shown with '-'. 
       Elements C and O are partly depleted into grain cores.
         The grain core elements are used only in  \ttfamily{RAMSES}\rm, where the core abundances are given by 
         $X[i_\mathrm{core}]=X[i_\mathrm{solar}]-X[i_\mathrm{gas}]$. 
       The choice of dust core abundances follows \citet{godard09,lesaffre13}. 
   The \ttfamily{ALCHEMIC}\rm~  code does not allow assumptions about the dust composition (shown with '-').}
\end{table}

The chemical network we designed represents the main species and reactions necessary to describe CO abundances 
within mainly H-C-O chemistry in the early phases of protostellar collapse \citep{lesaffre05}.
We include 14 neutral species (H, H$_2$, He, C, CH$_{x}$ with $x=1,...,4$, O, O$_2$, H$_2$O, OH, CO, and CO$_2$), their corresponding single-charged positive ions, 
and ionized molecules CH$_5^+$, H$_3$O$^+$, HCO$^+$, H$_3^+$. In addition, iron is also included as a representative metal. We choose Fe because it has lower sublimation temperature than other metal-type species \citep[i.e. see][and complementary tables at  
\href{http://www.astro.cornell.edu/~rgarrod/resources/}{www.astro.cornell.edu/$\sim$rgarrod/resources/}]{belloche14}.
A similar reduced network was used in \citet{wiebe03} for the case of molecular clouds, showing that one can calculate the abundance of carbon monoxide and fractional ionisation accurately with significantly reduced chemical networks in the case of pure gas-phase chemistry.  

We assume the solar total elemental abundance \citep{anders89,godard09}, of which the major part of metals and a part of oxygen and carbon are depleted onto the grains \citep[see Table~\ref{tab:param2} and][]{godard09,lesaffre13}.
The dust grains contain O, Si, Mg, Fe and C  as dust core  elements.
The size and mass of dust grains are calculated 
from the total mass of ``dust-core'' and ``dust-mantle'' elements and from the
adopted size distribution. 
The number of dust grains are calculated as $n_{\rm dust}=M_\mathrm{G,total}/ ((4/3)\pi \rho_{\rm solid}<a>^3$), where $\rho_{\rm solid}$ is the internal density, $M_\mathrm{G,total}$ is the total mass of the dust core elements, and $<a>$ is the mean radius calculated using  MRN size distribution $n(a) \propto a^{-3.5}$  \citep[][see appendix \ref{sec:ADust}]{mathis77}. 
Note that all dust core species  do not participate in the chemical reactions and are used exclusively to calculate dust grain mass, because we neglect the core erosion processes in the frame of the presented models. 
 All neutral species are allowed 
to freeze out on the dust grains, thus forming the grain mantles and increasing the weight of the grains. 
   Last, we consider three type of grains, namely neutral (G), single-charged positive (G$^+$), and single-charged negative (G$^-$). Table~\ref{tab:param1} gives a summary of all the chemical species we include in our reduced chemical network. 
In total, the reduced chemical network describes the evolution of $N_\mathrm{species}=56$ species.

%
\begin{table*}[t]
\center
\begin{tabular}{ccccccccccccccc}
\hline
\hline
\multicolumn{15}{c}{ Neutral species}\\
\vspace*{3mm}
$\rm H$ & $\rm H_2$ & $\rm He$ & $\rm C$ & $\rm CH$ & $\rm CH_2$ & $\rm CH_3$ & $\rm CH_4$ 
& $\rm O$ & $\rm O_2$ & $\rm OH$ & $\rm H_2O$ & $\rm CO$ & $\rm CO_2$ & $\rm Fe$ \\
\hline
\multicolumn{15}{c}{ Ionized species}\\
\vspace*{3mm}
$\rm H^+$ & $\rm H_2^+$ & $\rm H_3^+$ & $\rm He^+$ & $\rm C^+$ & $\rm CH^+$ & $\rm CH_2^+$ & $\rm CH_3^+$ & $\rm CH_4^+$ & $\rm CH_5^+$ & $\rm O^+$ & $\rm O_2^+$ & $\rm OH^+$ & $\rm H_2O^+$ & $\rm H_3O^+$ \\ 
 $\rm CO^+$ & $\rm HCO^+$ & $\rm Fe^+$ & & & & & & & & & & & & \\
\hline
\multicolumn{15}{c}{ Core species}\\
\vspace*{3mm}
 $\rm O^{**}$ & $\rm Si^{**}$ & $\rm Mg^{**}$ & $\rm Fe^{**}$ & $\rm C^{**}$ & & & & & & & & & &  \\
\hline
\multicolumn{15}{c}{ Mantle species }\\
\vspace*{3mm}
$\rm H^{*}$ & $\rm H_2^{*}$ & $\rm He^{*}$ & $\rm C^{*}$ & $\rm CH^{*}$ & $\rm CH_2^{*}$ & $\rm CH_3^{*}$ & $\rm CH_4^{*}$ 
& $\rm O^{*}$ & $\rm O_2^{*}$ & $\rm OH^{*}$ & $\rm H_2O^{*}$ & $\rm CO^{*}$ & $\rm CO_2^{*}$ & $\rm Fe^{*}$ \\
\hline
\multicolumn{15}{c}{ Grains}\\
\vspace*{3mm}
 $\rm G$  & $\rm G^+$ & $\rm G^-$ & & & & & & & & & & & & \\
\hline
\end{tabular}
\caption{ \label{tab:param1}  List of the chemical species included in the reduced chemical network. }
\end{table*}

\subsubsection{Chemical reactions \label{chemreact}}

We include gas-phase, freeze-out, and sublimation reactions. We do not take into account other types of reactions such as soft X-ray ionisation which is important for low density \citep[visual extinction A$_\mathrm{v}\sim0.2$,][]{wolf95}. For protostellar collapse applications, A$_\mathrm{v}$ is already larger than 10 for a typical mass of 1 M$_\odot$.

In total, our reduced gas-grain chemical network for H-C-O (and Fe as a representative metal) includes 231 reactions.

\paragraph{Gas-phase reactions.}

We include only reactions with two reactants. 
We use the classical Arrhenius representation for the reaction rate,
\begin{equation}
k_\mathrm{two}(T)=\alpha \left(\frac{T}{300~\mathrm{K}}\right)^{\beta} \exp{\left( -\frac{\gamma}{T} \right)},
\end{equation}
where $\alpha$ is the reaction rate at the room temperature,  $\beta$ characterizes the dependence on
the gas temperature, and $\gamma$ is the activation barrier. Most of the reactions constants we use have been downloaded from the online database KIDA \footnote{kida.obs.u-bordeaux1.fr} \citep[][see Sec.~\ref{sec:benchmark}]{wak12}.

\paragraph{Recombination. }
Ions can be divided in two groups: the ones which recombine at relatively low rate (slow) with free electrons (Fe$^+$, C$^+$, He$^+$, and H$^+$), and other ions (see Table~\ref{tab:param1}) which have up to three orders of magnitude higher recombination rate. To keep the network compact, we include the recombination reactions of all 'slow' ions and selected 'fast' ones: H$_3^+$, CH$_{3}^+$, CH$_{4}^+$, CH$_{5}^+$, O$_2^+$, H$_3$O$^+$, and HCO$^+$. Similarly, the charge exchange between ions and neutral Fe takes place only with the most abundant ions: H$^+$, C$^+$, H$_3^+$, O$_2^+$, H$_3$O$^+$, and HCO$^+$.  
Charge transfer is also occurring via dust grains, for all ions, electrons and types of grains.

\paragraph{Photodissociation and ionization. }
The envelope of collapsing low-mass dense molecular cores is usually dense enough so that we can safely assume a visual extinction from the exterior of about $A\mathrm{v} \simeq 10$ as previously mentioned. 
The rate is calculated as 
\begin{equation}
k=\alpha \exp{\left( -\gamma A_\mathrm{v} \right)}G_0 S,
\end{equation}
where $G_0$ is the far-UV (FUV) flux in units of the FUV interstellar radiation field of \citet{Draine78}.
The dissociation of $\rm H_2$ and $\rm CO$ are calculated with mutual self-shielding,
where $S$ is a shielding factor which depends on the column densities of $\rm H_2$ and $\rm CO$.
These shielding factors are necessary because the photodissociation of  $\rm H_2$ and $\rm CO$ occurs
only through discrete absorption lines. The $S$-factor for $\rm H_2$ depends on $\rm H_2$ column density,
whereas the shielding factor for $\rm CO$ depends on both $\rm H_2$ and $\rm CO$ column densities, because
some CO lines are shielded by $\rm H_2$ lines. The column density is calculated as a sum of external and internal column densities. The external column density is a parameter to be chosen in initial conditions 
(see paragraph \ref{init-cond}) and describes the properties of gas outside of the computational box, 
i.e. how deep the collapsing cloud is embedded into molecular clouds.
The internal column density has to be calculated inside the collapsing cores, as described in \citet{valdivia14}.

\paragraph{Gas-grain interaction. }

Dust properties (size, temperature, emissivity) are found to vary from one line of sight to another among
diffuse and dense clouds \citep{ste10}. These observations indicate that dust grains evolve throughout the interstellar medium. 
They can grow by the formation of refractory or ice mantles, or by coagulation into aggregates in dense and 
turbulent regions. They can also be destroyed by fragmentation and erosion of their mantles under more
violent conditions. 
Regarding the coreshine effect \citep{Pagani10}, the Spitzer L183 images can be best reproduced, if the 
dust grows starting from an 'initial' mean radius 
$a_0=0.05~\mu$m in the outer regions of dense clumps \citep[for density $n\leq{} 3\times{}10^{4}\rm ~cm^{-3}$, ][]{ste10}.
 For higher densities, the mean dust radius increases proportionally to gas density. 
The dust growth models \citep{ossen93,orm09} 
show that dust can grow by forming fractal aggregates, which are then compacted, if the 
dark clumps are older than several of $10^6$ years. 
Recent works \citep{Andersen13L260, Steinacker14L1506C,Lefevre14multi} demonstrate that dust growth can also explain the observed coreshine in several objects, even if they are younger than $10^6$ yrs.
We do not consider dust growth here and adopt a fixed distribution of dust sizes. Keeping in mind the dust growth issue, we vary the dust size as a parameter to determine its importance for chemical abundances (see Table~\ref{tab:collap}).

The neutral-dust reactions include freeze-out and desorption of species on(from) dust mantles. 
Dust can also interact with ions and electrons in charge-transfer reactions, which lead to neutralization of species.
The freeze-out reaction rate is 
\begin{equation}
k=\gamma \pi a^2 \sqrt{\frac{2k_\mathrm{B}T}{m_\mathrm{dust}}} \rm ~s^{-1},  
\label{eq:freezeout}
\end{equation}
 where $m_{\rm dust}=4/3 \pi a^3 \rho_{\rm solid}$ with $\rho_{\rm solid}$ for internal density of dust grains. 
The thermal desorption rate is
 \begin{equation}
k=q_{\nu} \exp{\left(-\frac{\beta}{T}\right)}  \rm ~s^{-1} , 
\label{eq:desorption}
\end{equation}
where $q_{\nu}$ is the vibrational frequency and $\beta$ is the critical temperature for thermal desorption of a specie.
The charge exchange reaction rate is given by
\begin{equation}
k=\gamma \left( \frac{T}{300 ~\mathrm{K}}\right)^{0.5} \left( 1+ \frac{450 ~\mathrm{K}}{T}  \right) \rm ~s^{-1}. 
\label{eq:chtransfer}
\end{equation}
We exclude the reactions describing the formation of complex organic molecules on grain surfaces.

The elaborated treatment of dust, linking the mass and size to its core composition, allow us to evolve dust via erosion reactions.  For collapse, it can become important when temperatures are rising above 800 K (chemo-erosion of C). In the hotter interior of the FHSC, the evaporation of Si$^{**}$ and $\rm Fe^{**}, \rm Mg^{**}$ can be accounted for. This will be used in future works. For the current study, we exclude dust-core erosion processes and keep the amount of solid material fixed (with the exception of ices, i.e. mantle species).

\paragraph{Cosmic ray (CR) ionization and desorption.}

We use a constant CR ionization rate $\zeta_{\rm CR}$ in this paper. In reality,
the impact ionisation by  cosmic rays decreases with increasing 
density of the gas, because the cosmic rays, which follow magnetic field lines, have to cross
a significant bulk of mass before reaching the dense regions in the inner part of the collapsing core \citep{pad11}. 
Cosmic rays may give rise to secondary electrons with energy of about 30 eV
 \citep{crav78}, whereas the photons with energies of around 10 eV  can ionize and 
dissociate species in the 
gas phase, as well as detach electrons from grains. 
These secondary electrons excite electronic states in H$_2$ via collisions,
 which result in ultraviolet 
fluorescence. The rates of photodissociation and photoionization are dependent on the relative
numbers of photons which are absorbed by the gas and by the
dust. 
To describe the rates for ionization, dissociation or desorption induced by CR and secondary UV photons, we use
\begin{equation}
k_{\rm CR}=\gamma \zeta_{\rm CR} \left(\frac{T}{300 ~\rm K}\right)^{\alpha},
\end{equation}
where $\gamma$ is the value of the reaction rate at the room temperature of 300 K, and
 $\alpha$ is  zero for all species but CO. This unusual dependence of CR ionization 
on gas temperature is a 'short-cut', allowing to take into account CO dissociation by CR-induced secondary photons \citep[see for details][]{gredel87,flower07}. 

We include 21 reactions of ionization, which contain 9 reactions for the ionization of H, He, H$_2$ (with two ways), O$_2$, C, CO, CH$_2$ and CH$_3$, as well as 12 reactions for dissociation
of H$_2$, CH, CH$_3$, CH$_4$ (two ways), CH$^+$ (two ways), H$_2$O, O$_2$, OH, CO and CO$_2$. 

Reaction rates for CR-desorption of the species from the dust surface is 
equivalent to thermal desorption (see eq.~\ref{eq:desorption}), 
under the assumption that CR heats the grain up to roughly 70~K \citep{semenov10}. 
We include 13 reactions since H$_2$ and He are released in the gas phase already at 10~K.

\subsubsection{The chemistry module}

We use the time-dependent chemistry {\ttfamily PDS} code, 
originally written for MHD shocks \citep{flower15,flower03,lesaff04,lesaffre05,lesaffre13}, 
which we have improved and successfully tested for physical conditions typical of the FHSC formation problem 
(see Sec.~\ref{sec:benchmark}). 
Modelling of the chemical processes is described via a set of equations in the form
\begin{equation}
\partial_t{n_x} =  - n_x \sum_r k_{rx}n_r + \sum_p \sum_q k_{pq}n_pn_q, \label{chem_solver}
\end{equation}
where $n_x$ is the number density of a $x$ chemical species. 
On the right hand, the negative term is the sum over the destruction reactions, and positive term is the sum
over the reactions creating  $x$-species from species $p$ and $q$. $k_{rx}$ and $k_{pq}$ are the
corresponding reaction rates.
The cooling via molecular and atomic lines is present in {\ttfamily PDS} chemical code.
In the context of this paper,
we neglect the thermal feedback of chemistry on the gas to concentrate on which parameters are affecting 
the coolant's abundances. 

We solve the system of equations  (\ref{chem_solver}) using either explicit or implicit integration in time. 
When the chemical time step is shorter than the hydrodynamical one, we use an implicit solver which consists in inverting the matrix 
of $N_\mathrm{species}^2$ in size for a hydro(-magnetic) timestep of $\Delta t_\mathrm{hydro}$. Should this not be the case, 
the chemical module switches to an explicit second-order Runge-Kutta method for time advancement.

In the implicit scheme, we use the DVODE solver\footnote{from the ODEPACK package downloadable at https://computation.llnl.gov/casc/odepack/}, without any optimization of the matrix inversion procedure.
There are indeed various optimization methods, like buffering, using the topology of matrix, flux method to remove slow reactions, or rates tabulation \citep{grassi12,grassi13}. 
The highest possible speed gains here can be achieved for large networks, when using
sparse representation of Jacobi matrix and sparse solvers of linear system of equations,
like it is done in {\ttfamily ALCHEMIC} code \citep{semenov10}.
According to our experience with the reduced network in {\ttfamily RAMSES},
 those optimization methods can rather slow down the performance and even lead to the wrong abundances of the species, when applied blindly. 

In this study, we present a set of models with non-optimized chemistry module, 
which shall serve as a benchmark for the upcoming optimized simulations. Typically,
non-optimized chemo-dynamical simulation costs about 100 times more computational time compared to purely RHD collapse calculations.

\paragraph{The convergence control.} 
The equations for chemical evolution are often stiff, i.e. having a large range of reaction time scales. 
In this case, the impicit solver will be used in the chemistry module. 
In this paragraph we briefly describe the solver and its convergence.
 
The chemistry solver proceeds in two steps: first,  the
hydrodynamical time-step $\Delta t_\mathrm{hydro}$ is divided into smaller chemical time-steps $\Delta t_\mathrm{chem}$. 
Within the call of the DVODE solver, $\Delta t_\mathrm{chem}$ can also be reduced again automatically if the solver recognizes a problem (see ODEPACK package for details).
The chemical solver stops when the sum of the intermediate $\Delta t_\mathrm{chem}$ reaches $\Delta t_\mathrm{hydro}$. 
In each call of the DVODE solver, 
the set of equation (\ref{chem_solver}) is integrated using Newton-Raphson iterations until convergence is obtained, which is controlled by
the DVODE tolerance parameters $\epsilon_\mathrm{tol}$ on the change of the absolute or relative abunancies ($\epsilon_\mathrm{tol}= 10^{-6}$ typically). 

\subsection{Benchmarking of the reduced chemical network \label{sec:benchmark}}

In this section, we compare our reduced chemical network performance with the results of the 
full gas-grain chemical network \ttfamily{ALCHEMIC}\rm~\citep{semenov10}.
We study the time evolution during $10^6$ years of the chemical species
for '0-D' case, assuming constant gas density and temperature, 
$n=10^4 {\rm ~cm^{-3}}$ and $T=10 {\rm ~K}$. Our benchmark tests include pure gas-phase chemistry case, 
the case when all neutral species freeze out, and the case where both freeze-out and thermal desorption are included. 
 We do not include the reactions responsible for the  water formation on the dust surface,  what saves the computational time. 
The limitation here is an under-estimation of the total water abundance during the collapse simulation.
The dust-surface reactions would lead to $\rm X[H_2O]>X[CO]$ on the long term, 
i.e. when the molecular cloud is older than 1 Myr. 
The water molecules should be the main reservoir for oxygen in older clouds. 
In younger objects,
 $\rm CO$ is the dominant reservoir for oxygen.

At the beginning of benchmark tests, all initial chemical abundances including ices are set to $10^{-19}$, except for $\rm H$, $\rm H_2$, $\rm O$, Fe, $\rm C^+$, and the elements in the dust cores for which we follow \citep[][see Table 1 and 2]{godard09}.
 The dust-to-gas ratio $f_\mathrm{dg}$ is assumed 0.01 in \ttfamily{ALCHEMIC}\rm, and calculated in  \ttfamily{RAMSES}\rm~ using $n_{\rm dust}/n_\mathrm{H}=3\times{}10^{-12}$. 
We force the properties of dust to be exactly the same in both codes, 
fixing the dust size to be 0.1 $\mu$m. 

There are some significant differences in the reaction coefficients which deserve to be justified.
 \ttfamily{ALCHEMIC}\rm~ uses the reaction coefficients from the KIDA online data base. 
The reason for these differences in reaction coefficients in our network is 
that the network has been reduced 'by hand', compensating the missing reactions 
so that CO and $\rm HCO^+$ abundances could be
obtained {\it very closely} to the abundances from full chemical network of \ttfamily{ALCHEMIC}\rm.   
Thus, few coefficients in the reduced network are in
fact {\it effective} coefficients.%

In comparison to the original version of {\ttfamily PDS} code \citep{flower15,lesaffre05,lesaffre13}, we modify some reactions
and partly update the reaction coefficients using the KIDA database. 
We have been then able to update 
the reaction involving   $\rm H^+$ and $\rm CH_4$ using KIDA coefficients to match closer match the  
 {\ttfamily ALCHEMIC}'s relative abundances of these species. 
We also update the treatment of 
cosmic ray and thermal desorption following the work of \citet{herbst08}.

To validate our reduced chemical network, we proceed in four steps with gradual increasing complexity in the chemistry. Intermediate results are presented in appendix~\ref{sec:benching}.  First, we compare the results of the two different networks for the case of pure gas-chemistry and achieve a very good  match except for CO$_2$ (fig.~\ref{abumol1}).
Then we add step-by-step adsorption on the grain surface (fig.~\ref{abumol2}), and
CR-desorption (fig.~\ref{abumol3}). 
We observe very satisfactory agreement of the chemical evolution for all tests.

\begin{figure*}
\begin{center}
\includegraphics[width=6.6in]{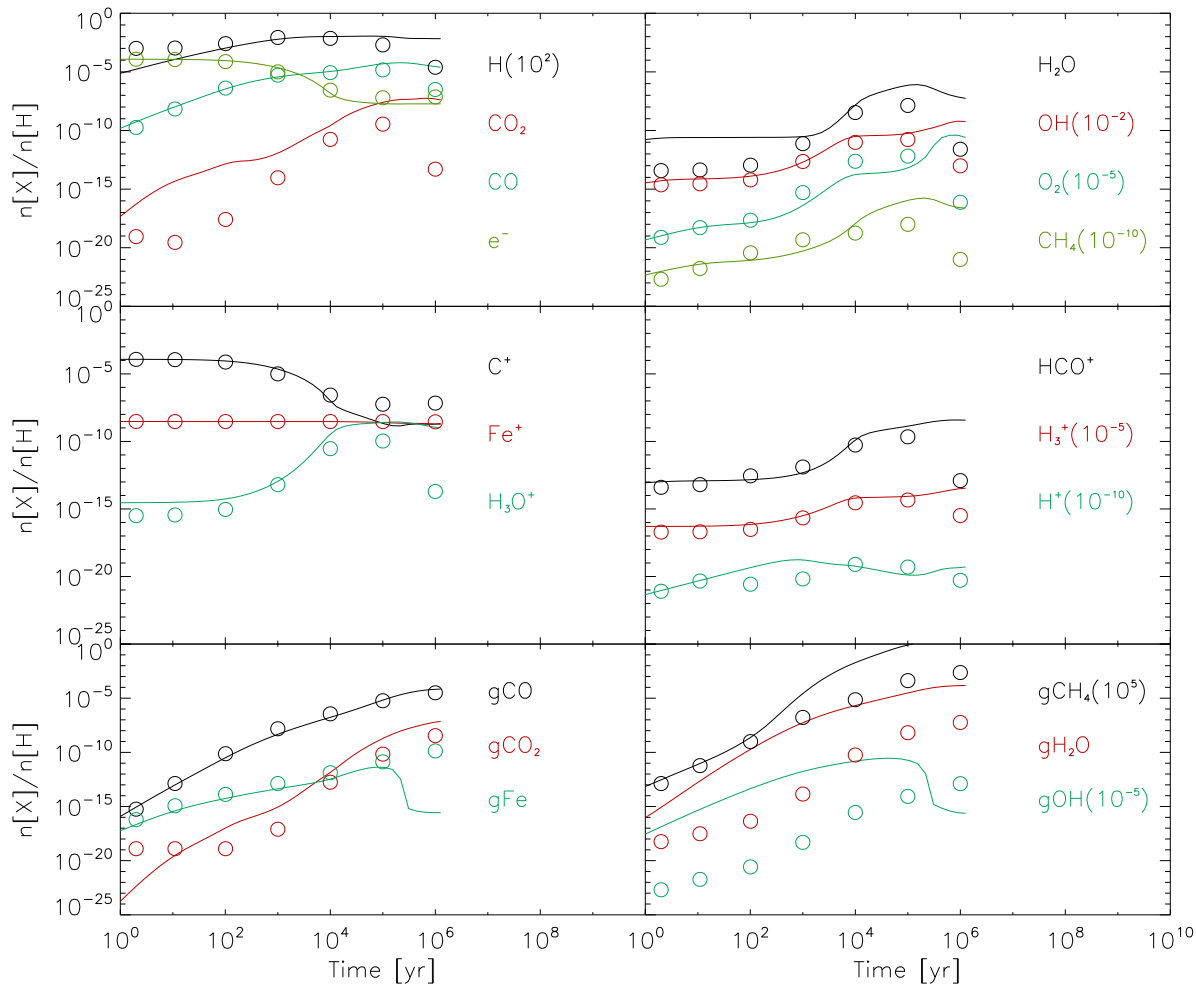}  
  \caption{ Time evolution of selected species for complete set of gas-phase and  
              gas-grain reactions (freeze-out on 
           grain surface, CR and thermal desorption reactions). We plot a relative chemical abundances defined as $n[X]/n[H]$,
           where $n[H]=\rho_{\rm gas}/m_{\rm p}$ and $m_{\rm p}$ is the mass of proton. 
           Solid line shows full chemical network {\ttfamily ALCHEMIC} \citep{semenov10}, 
            circles show reduced network with {\ttfamily RAMSES and PDS} codes. 
      Note that the full chemical network {\ttfamily ALCHEMIC} includes additional reactions 
            between ices on the dust surface. 
            }
\label{abumol4}
\end{center}
\end{figure*}

As the 
last step the thermal desorption is included in {\ttfamily RAMSES} code, and compared with full strength of {\ttfamily ALCHEMIC} code which includes also the formation of complex organics on the dust grains (fig.~\ref{abumol4}).
The differences we observe in fig.~\ref{abumol4} teach us when those reactions on the dust surface start to play a crucial role. 
Our conclusion is that our reduced network performs well for dust-free chemistry, 
and in the presence of dust on the timescale shorter than $10^6-10^7$ years.    
On the longer timescales, there are severe differences in the temporal evolution of the chemical abundances  which are caused by missing species and reactions.
For example, we have too much of iron in the ice form  because we neglect FeH formation on the dust surface.
The amount of water and CO$_2$ in ice form is lower compared to {\ttfamily ALCHEMIC}'s abundances,  where the dust surface formation of water and other molecules is included. 
As our collapse is going to happen on much shorter time scale {(free-fall timescale $\sim10^4-10^5$ years)}, we suggest that the reduced network is fairly good for the simulations of dynamical chemistry during the dense cloud collapse.

%


\subsection{The chemo-radiation-magneto-hydrodynamic model \label{crmhd}}

To perform 3D time-dependent chemical calculations of protostellar collapse, we coupled the time-dependent chemistry code {\ttfamily PDS} code  \citep{flower15,flower03,lesaffre05,lesaffre13} with the 
{\ttfamily RAMSES}\rm\ code \citep{teys02}. 

\subsubsection{The AMR RMHD code {\ttfamily RAMSES}\rm}
 We use the radiation-magnetohydrodynamics (RMHD) solver of {\ttfamily RAMSES}. The MHD part of the code is based on the Constrained Transport (CT) scheme \citep{teys06}, 
using a 2D-Riemann solver to compute the electromotive force at cell edges \citep{From06,teys06}.
In addition to the ideal MHD solver, ambipolar diffusion and Ohmic diffusion are also implemented as additional electromotive forces in the
induction equation \citep{mass12}.  For radiation transfer, we use the grey flux-limited diffusion (FLD) approximation 
described in detail in \citet{comm11a}, which combines the explicit second-order Godunov solver of {\ttfamily RAMSES}\rm\ for the hydrodynamical part, and an implicit scheme for radiative energy diffusion and coupling between matter and radiation 
terms. The implicit FLD solver uses  adaptive time stepping \citep[ATS, ][]{comm14} which enables us to benefit from the ATS scheme designed for the hydrodynamical part in {\ttfamily RAMSES}\citep{teys02}. The gain in computational time over the former unique time step method ranges from 
5 to 50 depending on level of adaptive time-stepping and on the problem.

\subsubsection{The chemo-RHD solver}
In this study, 
we neglect the effect of magnetic fields to concentrate on the dynamical evolution of chemistry coupled 
to an accurate description of the thermal balance of the gas during the collapse.
The set of chemo-RHD equations solved in \ttfamily{RAMSES}\rm \ with all radiative quantities estimated in the co-moving frame are
 \begin{equation}
\left\{
\begin{array}{llll}
\partial_t \rho + \nabla \cdot\left[\rho\vec{u} \right] & = & 0 \\
\partial_t \rho \vec{u} + \nabla \cdot \left[\rho \vec{u}\otimes \vec{u} + P \mathbb{I} \right]& =& - \lambda\nabla E_\mathrm{r} \\
\partial_t E_\mathrm{T} + \nabla \cdot \left[\vec{u}\left( E_\mathrm{T} + P_\mathrm{} \right)\right] &= & - \mathbb{P}_\mathrm{r}\nabla\cdot{}\vec{u}  - \lambda \vec{u} \nabla E_\mathrm{r} \\
 & & +  \nabla \cdot\left(\frac{c\lambda}{\rho \kappa_\mathrm{R}} \nabla E_\mathrm{r}\right) \\
\partial_t E_\mathrm{r} + \nabla \cdot \left[\vec{u}E_\mathrm{r}\right]
&=& 
- \mathbb{P}_\mathrm{r}\nabla\cdot{}\vec{u}  +  \nabla \cdot\left(\frac{c\lambda}{\rho \kappa_\mathrm{R}} \nabla E_\mathrm{r}\right) \\
 & &  + \kappa_\mathrm{P}\rho c(a_\mathrm{R}T^4 - E_\mathrm{r})\\
 \partial_t{n_x}/+\nabla \cdot [n_x \vec{u}] & = & - n_x \sum_r k_{rx}n_r  \\
  & & +  \sum_p \sum_q k_{pq}n_pn_q,
\end{array}
\right.
\end{equation}
\noindent where $\rho$ is the material density, $\vec{u}$ the velocity, $P$ the thermal pressure, $\kappa_
\mathrm{R}$ the Rosseland mean opacity,  $\lambda$ the radiative flux limiter \citep[e.g.][]{min78}, $E_\mathrm{T}$ the total energy $E_\mathrm{T}=\rho\epsilon +1/2\rho u^2 + E_\mathrm{r}$ ($\epsilon$ is the internal specific energy), $\kappa_\mathrm{P}$ the Planck opacity, $E_\mathrm{r}$ the radiative energy, and $\mathbb{P}_\mathrm{r}$ the radiation pressure. Note that compared to classical RHD solvers, we now have additional $N_\mathrm{species}$ equations on the number density of each specie $n_x$, which correspond to the advection, chemical creation and destruction of the $x$ chemical elements.

Currently, the {\ttfamily PDS} chemical code has no impact on the evolution of the gas in RHD part of {\ttfamily RAMSES}. 
The possible limitations are then three-folds. First, the total gas and dust density $\rho$ is not recomputed from the abundances obtained by the chemistry solver. 
This is not affecting our results, because we consider FHSC before the dust melting occurs above 800~K. Second, as mentioned previously, we do not include heating and cooling due to atomic and molecular lines, which should not be dominant as long as radiative transfer is dominated by the dust. Cooling by CO lines can be expected to become important for the gas densities of $10^3 \rm cm^{-3}$ and lower. 
Last, we do not compute self-consistently the dust opacity from the dust composition given in output by the chemistry solver.
 For the latter, we instead use tabulated opacities from \citet{sem03}. 
All these limitations could be dealt by the combined {\ttfamily RAMSES} and {\ttfamily PDS} codes, but go far beyond the scope of the paper.

\subsubsection{Requirement of element conservation. }

The most strict and model-independent input for our chemistry code
is the declaration of total elemental abundances, which we take equal to Solar abundances.
They include oxygen, carbon, nitrogen, sulfur, and metals such as
silicon, magnesium and iron. The maximal size of the chemical network include species involving 
those elements. 
Any time-dependent chemo-dynamical solver must guarantee total element abundances conservation {\it for each time step}, 
which can suffer from various numerical effects. 
First, element conservation is required during the chemical evolution. 
Second, the conservation has to be enforced during the advection.

The ``consistency'' subroutine is constructed so that after each advection operation the local 
elemental abundances are conserved (Modified Consistent Multi-fluid Advection). 
As pointed out in works by \citet{glov10}, this is essential for correct evoluation of chemistry.

The ``consistency'' check also takes into account the composition of dust, or the composition changes.
In the case of sputtering of the species from the  dust mantles or erosion of the dust core, 
the total amount of chemical elements in both gaseous and solid 
form is conserved.

\section{Results for dense core collapse calculations}

\subsection{Initial conditions \label{sec:condinit}}

\begin{table*}[t]
\vspace*{3mm}
\begin{tabular}{lllllllll}
\hline
\hline
\vspace*{3mm}
Model & $E_{\rm th}/E_{\rm grav}$ & $r_{0}$[pc]   & $\sqrt{<a^2>}[\mu$m] & $n_\mathrm{dust}/n[{\rm H}]$ & 
$n_{\rm dust}<a^2>/n[{\rm H}]$ & $t_\mathrm{0}$ [kyr] & section\\
\hline
S1 & 0.45  & 0.022  & 0.1 & 3.1$\times{}10^{-12}$ & 3.1$\times{}10^{-14}$ &  57.8  & Sec.~\ref{sec:fidu} \\ 
S1x2 & 0.89 & 0.045    & 0.1 & 3.1$\times{}10^{-12}$ & 1.0$\times{}10^{-14}$  &  455.0  & Sec.~\ref{sec:freefall} \\ 
S1x4 & 1.79 & 0.088   & 0.1 & 3.1$\times{}10^{-12}$ & 1.0$\times{}10^{-14}$ &  1702.0  & Sec.~\ref{sec:freefall} \\ 
\hline
S2$_{\rm MRN}$ & 0.45 & 0.022   & 0.05 & 3.9$\times{}10^{-12}$ & 1.0$\times{}10^{-14}$  & 57.8  & Sec.~\ref{sec:dust}   \\
S3$_{\rm MRN}$ & 0.45 & 0.022   & 0.017 & 5.2$\times{}10^{-11}$ & 1.5$\times{}10^{-14}$ & 57.8 & Sec.~\ref{sec:dust}   \\
S4 & 0.45 & 0.022   & 1.00 & 3.1$\times{}10^{-15}$   & 3.1$\times{}10^{-15}$& 57.8  & Sec.~\ref{sec:dust}   \\
\hline
\end{tabular}
\caption{ \label{tab:collap}  List of 3D chemo-dynamical calculations of collapse. 
   The cloud is assumed to be always of one solar mass. S1 is a fiducial model. 
   Model S1x2 and S1x4 are done for the collapse of the cloud of same mass but spreaded over larger calculation box $L_{box}$, i.e. core radius $r_0=L_{\rm box}/4$ is increasing factor 2 and 4, and so the relation between thermal and gravitation energy is altered.
   S2$_{\rm MRN}$, S3$_{\rm MRN}$  are the models with MRN dust size distribution, whereas
    models S1, S4 have fixed grain size. $n_{\rm dust}a^2$ represents mean dust cross-section. 
    }
\end{table*}

\paragraph{Parameters of the core}
We choose a spherically-symmetric collapse
configuration, i.e. we neglect rotation, magnetic fields and turbulence. 
The initial core mass is fixed to 1 M$_\odot$ and the temperature of both gas and dust is uniform and equals 10 K. 
Note that in our model, we assume that the dust and gas are perfectly coupled thermally. 
The adiabatic index  is $\gamma=5/3$ and the mean molecular weight is $\mu_\mathrm{gas}=2.375$. 
In this paper we use both $n_{\rm gas}=\rho_{\rm gas}/(\mu_{\rm gas}m_p)$ and $n[{\rm H}]=n_{\rm gas}\cdot{}\mu_{gas}$, where $m_p$ is the mass of proton. It is common to represent the relative abundance of $x$ chemical specie as $n[x]/n[{\rm H}]$.
In all our models, the initial density profile is Bonnor-Ebert like $n= n_\mathrm{c}/(1 + (r/r_\mathrm{c}))^{-2}$,
where the maximum density in the center $n_\mathrm{c}$ is 10 times larger than the density at the core's border. 
The total length of the simulation box is four times larger than the core initial radius $r_0$. 

Table \ref{tab:collap} gives a summary of all the calculations we present in this study. 
In our fiducial model, labeled S1, the central density is $n_\mathrm{c}=4.4 \times 10^{5}$ cm$^{-3}$ (or, $1.71\times10^{-18 }$ g cm$^{-3}$), and the core radius is $r_0=0.022$ pc. 
The relation between thermal and gravitational energies is $E_{\rm th}/E_{\rm grav}=0.447$.

First, we add two more models, which have the initial core mass but different box length, i.e a factor two (S1x2) and four (S1x4) larger. The initial density and thus the free-fall time t$_\mathrm{ff}$ thus increases as the central density decreases.  This affects the time available for the 
chemical reactions to evolve, especially affecting the abundances of ices (see Table~\ref{tab:collap}).
Second, we take the same core initial properties as model S1, but we vary the dust properties (models S2 to S5). 

\paragraph{Choosing/generating initial chemical abundances \label{init-cond}}
Defining a good initial chemical abundance for a dense core is a topic in itself.
The physics of dense core formation is not yet well understood. 
What is the lifetime of molecular cloud prior to collapse,  
how fast do dense core collapse, what happens to dust size distribution, what is the cosmic ray ionisation rate, 
 the exposition to X-rays and UV photons? There are no precise answers to those questions. 
To simplify and focus on the effect of dynamical chemistry alone, we choose a uniform high visual extinction $A_\mathrm{v}=30$ and a uniform, cosmic ray ionization rate of $1.3\times{}10^{-17}$ s$^{-1}$.
 In order to generate the initial
chemical abundances, we start from elemental abundances listed in Table~\ref{tab:param2} 
and let the 
chemistry evolve for $6\times{}10^5$ years in the static Bonner-Ebert sphere configuration  \citep[e.g.][]{hin13}. As mentioned previously, the assumptions about dust size distribution differs depending on the model (see Table~\ref{tab:collap}).
In the prevous benchmarking results, we have tested the reduced network for $n_{\rm gas}=10^4 \rm cm^{-3}$ and $T=10$K. We adopt similar value for the gas density in the fiducial model, with the central gas density order of magnitude higher. 
 It can be easily shown, that the relative abundances of CO and gCO are very insensitive to the gas density if temperature remains of about 10K. The ratio of $n[{\rm CO+gCO}]/n[{\rm H}]$ remains about $10^{-4}$ for the whole range of the gas densities from $10^4 \rm cm^{-3}$ to $10^{12} \rm cm^{-3}$, and is only limited by the availability of oxygen to form CO (see appendix~\ref{sec:cube}). 

\paragraph{Synchronization}
The properties of the FHSC are independent of the cloud mass and weakly affected by the temperature of the parental cloud
 \citep{vaytet12}. When we simulate the collapse of dense cores identical in all aspects except the initial density, the resulting FHSCs shall appear similar. Of course, each simulation will need different time to reach the FHSC depending on how dense it was initially, what may affect the chemical abundances.
Thus, we define here as a zero-time $t_0$ the moment of the FHSC formation, i.e. when central density reaches $10^{13}$ cm$^{-3}$ ($T\sim 210$ K). The zero-time for different simulations is summarized in Table~\ref{tab:collap}.

\subsection{Fiducial model: evolution of chemical abundances during the collapse \label{sec:fidu}}

\begin{figure*}
\begin{center}
\includegraphics[width=7.5in]{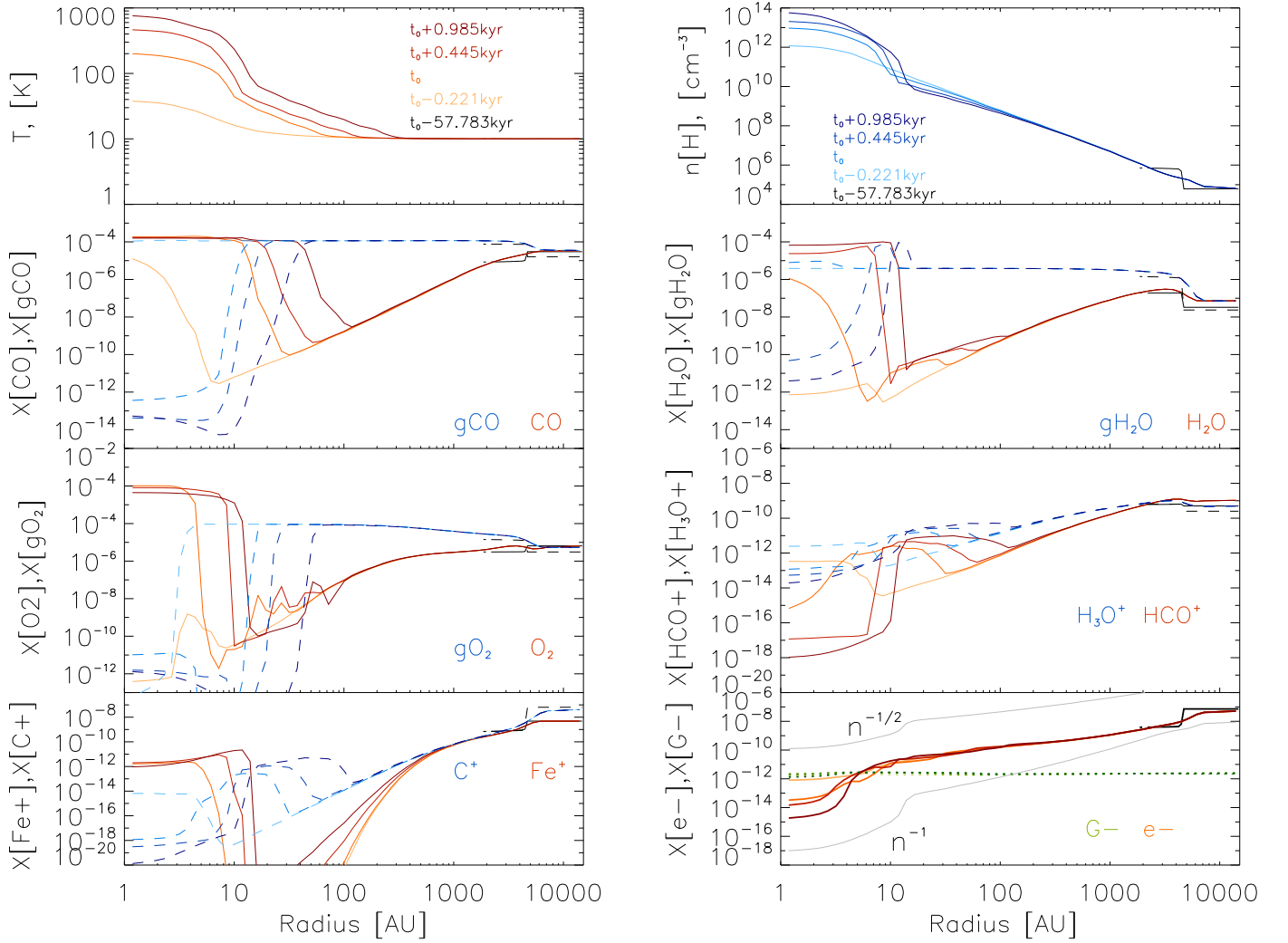}
  \caption{ Model S1: fiducial collapse model and corresponding time evolution of chosen species.}
\label{col1}
\end{center}
\end{figure*}

Figure~\ref{col1}~ shows the radial profiles for the fiducial run S1 of total density, temperature, and selected relative chemical abundances for five different times, from the start  at $-57.8$ kyr and up to the moment when the central temperature exceeds 800 K at $0.99$ kyr. 
At the beginning of the simulation, most of the neutral species are in the form of ice. After the FHSC formation, i.e the beginning of the adiabatic phase, thermal desorption puts the neutrals back to gas phase. This ice sublimation starts inside at the edge of the future FHSC (8 to 10 AU) and proceeds outwards with time as the FHSC accretes mass and radiate away all the accretion luminosity \citep{comm11b}. The chemically-related ion abundances develop a flat-shaped local maximum at roughly the same location where related neutrals are evaporated from the dust surface.  
The pairs of $(\rm CO-HCO^+)$, $(\rm H_2O-H_3O^+)$, and $(\rm CH_4-C^+)$ show a correlation between evaporation of
 the molecule and a bump in the ion abundances. 
 
In the bottom-right panel, we show the relative electron abundance profiles accompanied by relative abundances of negatively charged grains.
For radii $r>10$ AU, the number of charged dust is small compared to main charge carries, so that $n_\mathrm{i} \approx n_\mathrm{e}$. The grey lines show the profiles $\propto n_{\rm gas}^{1/2}$ and $\propto n_{\rm gas}^{-1}$ for comparison .
Profile $n_\mathrm{e} \propto n_{\rm gas}^{-1/2}$   originates from the early works on equilibrium ionization calculations \citep{elmegreen79,fiedler93}.
This fit is only valid when the gas density is smaller than $10^{10}$ cm$^{-3}$, i.e.  where the role of dust as charge carrier is relatively weak. In fig.~\ref{col1} we observe that indeed the electron abundance scales as  $n_{\rm gas}^{-1/2}$ for $r>10$ AU. 
In \citet{machida07}, the electron abundance scales as $n_{\rm gas}^{-1}$ and represents the scaling where charged dust is important. 
\citet{machida07} have considered much smaller dust size (distribution of dust grains with $a_\mathrm{min}=0.005 \mu $m and $a_\mathrm{max}=0.25\mu$m), what results in ion-dust dominated regime. 
Closer to the FHSC border, our electron abundances resemble  $n_{\rm gas}^{-1}$ because of the increased role of negatively charged dust. The  implications of ion-electron and ion-dust regimes for the magnetic diffusivities will be discussed in the subsequent paper. 

\begin{figure*}
\begin{center}
\includegraphics[width=6.6in]{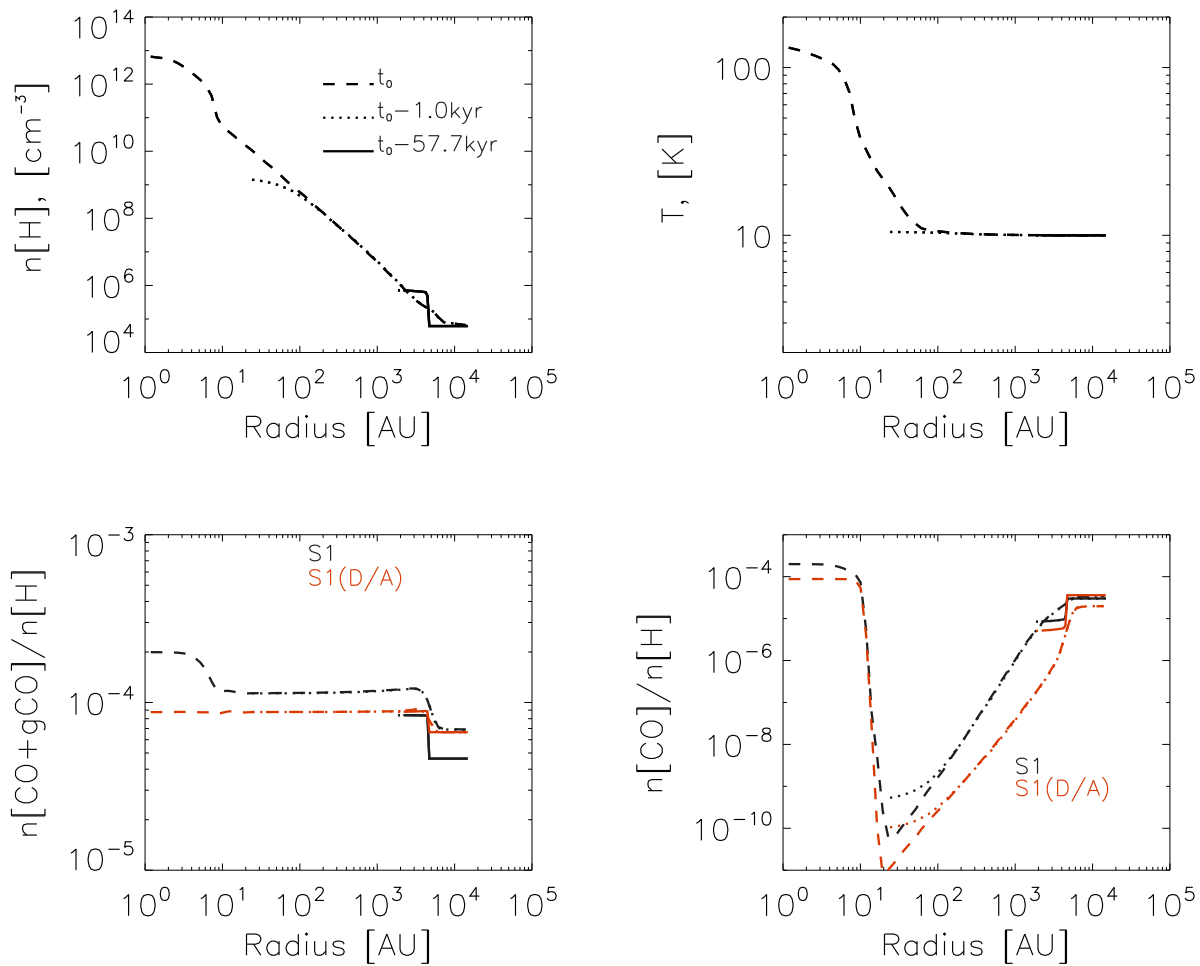}
  \caption{ Effect of dynamical chemistry on CO abundances. Top row: gas density and temperature radial profiles at three different times, shown with solid, dotted and dashed lines. 
 Left bottom: Total (gas phase and ice) CO abundance profiles for the fiducial model S1 in black, and for the same collapse with advection of species  but with only reactions of adsorption and desorption (model S1(D/A)), shown in red.
  Right bottom: fraction of gaseous CO for S1 and S1(D/A) models, same line and color coding.}
\label{follow0}
\end{center}
\end{figure*}

We are now interested in the seemingly flat profile observed for the ratio $(n[\mathrm{CO}]+n[\mathrm{gCO}])/n_{\rm gas}$ with radius. 
The isothermal phase of collapse takes most of the time, starting from $t_{0}-57.8 $ kyr (start of the simulation) to roughly $t_{0}-1$ kyrs. During this time, the abundances of CO and other molecules increase due to gas-phase reactions. 
We make an experiment to test whether the CO abundances can be reproduced with static chemical model, at least during the isothermal stage of collapse. 
We change the initial condition so that the chemical abundances in the initial cloud are $56.8$ kyrs older than in S1 model, matching the age of the output with $t-1$kyr. Then, we run a simulation with only  adsorption and desorption on(from) the dust grain surface (model S1(D/A)). The species are thus just advected and can only freeze or melt on/from the dust surface \citep[chemisorption of CO, see also][]{lesaffre05}. 
Figure \ref{follow0} shows only a small difference between models with chemically evolved total CO abundances (S1) and the pre-evolved total CO in case of S1(D/A), if the temperature remains close to 10 K (\ref{follow0} bottom right, fig.~\ref{follow1} as well). The difference remains small in spite of significant increase in the gas densities close to the FHSC border, which is in agreement with the results of the evolution map of CO shown in Appendix~\ref{sec:cube}. This is also in agreement with the finding of \citet{lesaffre05} and adds a further justification for using the post-processing techniques to account for the {\it total} CO abundances.
 
When plotting the ratio of gaseous CO to its total abundance (fig.~\ref{follow0}, bottom right), we can observe that stationary chemistry in S1(D/A) case underestimate the CO in the gas phase by more than one order of magnitude at radii $10^2 <r <4 \times10^3$AU. Gaseous CO is a tiny fraction of the total carbon monoxide abundance and this difference is invisible in fig.~\ref{follow0} bottom left. Thus, we note that post-processing with static chemistry may not be accurate enough to reproduce the abundance of CO in the gas phase.

After the temperature begin to rise within the FHSC, the total CO abundance increases by a factor of a few due to the 
faster reactions in the hotter gas. 
Above 60 K, the group of $\rm CH_x$ molecules is thermally desorbed from the grain surface. 
Hence, the abundance of CO increases by a factor of a few through the reactions  
\begin{equation}
\rm HCO^+ + CH_3 \to{} CH_4^+ + CO,\nonumber
\end{equation}
and
\begin{equation}
\rm HCO^+ + CH_4 \to{} CH_5^+ + CO.\nonumber
\end{equation}
The steady-state chemisorption of CO  will not apply inside of the core, or in regions where $T>60$ K which can extend further in the disk and in the outflow found in rotating collapsing cores \citep[e.g.][]{com10}.

%

\subsection{Impact of free-fall time on the chemical abundances in the FHSC \label{sec:freefall}}

\begin{figure*}
\begin{center}
\includegraphics[width=7.5in]{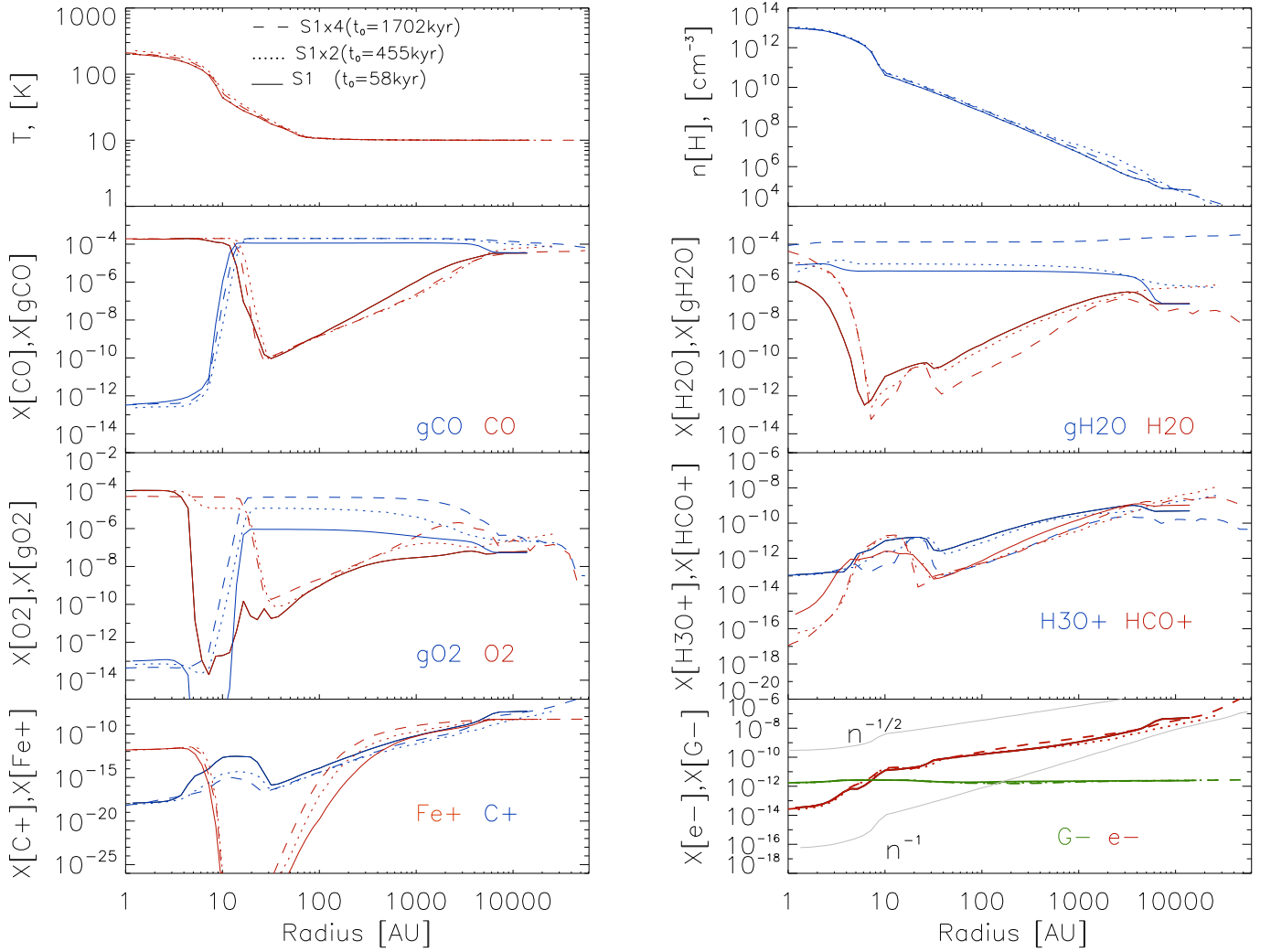}
  \caption{ Effect of free-fall time. 
            Solid line shows fiducial model (S1), dotted line the model with twice as large box (S1x2), and the
            dashed line the model S1x4 with a box four times larger. 
            In the top left window we provide the time $t$ elapsed 
            from the begin of simulation to the moment we choose as $t_0$.}
\label{col2}
\end{center}
\end{figure*}

In this section, we investigate the effect of free-fall time on chemistry, i.e 
the time for the chemistry to evolve prior to FHSC formation. 
Simulations runs S1x2 and S1x4 start with the same initial mass as fiducial model S1, 
but with a core radius two and four times larger so that the core's free-fall time 
increases as the initial density decreases (Table~\ref{tab:collap}).
Figure \ref{col2} presents a direct comparison of selected chemical abundances at $t_0$ as in fig.~\ref{col1}. 
There is strikingly more ice available at $t_0$ in model with largest initial core radius, S1x4. 
Water ice is even reaching $ n[\mathrm{gH_2O}]/n[\mathrm{gCO}]>1$ in the low-density region. 
Note that, as mentioned in Sec.~\ref{sec:benchmark}, we deliberately underestimate the amount of water ice due to the lack of the dust-surface reactions in our reduced chemical network. 

We can observe the features which are common with the fiducial model. 
Again, the pairs of $(\rm CO-HCO^+)$, $(\rm H_2O-H_3O^+)$ and $(\rm CH_4-C^+)$ show correlation between evaporation of the molecule and the bump in the ion abundances. 
The abundance of CO is increasing only by a factor of a few with the age of the cloud.   
The total amount of CO seems to be a fixed fraction of gas density everywhere, which is partly an effect of the log-scale of the plots. 

\begin{figure*}
\begin{center}
\includegraphics[width=7.in]{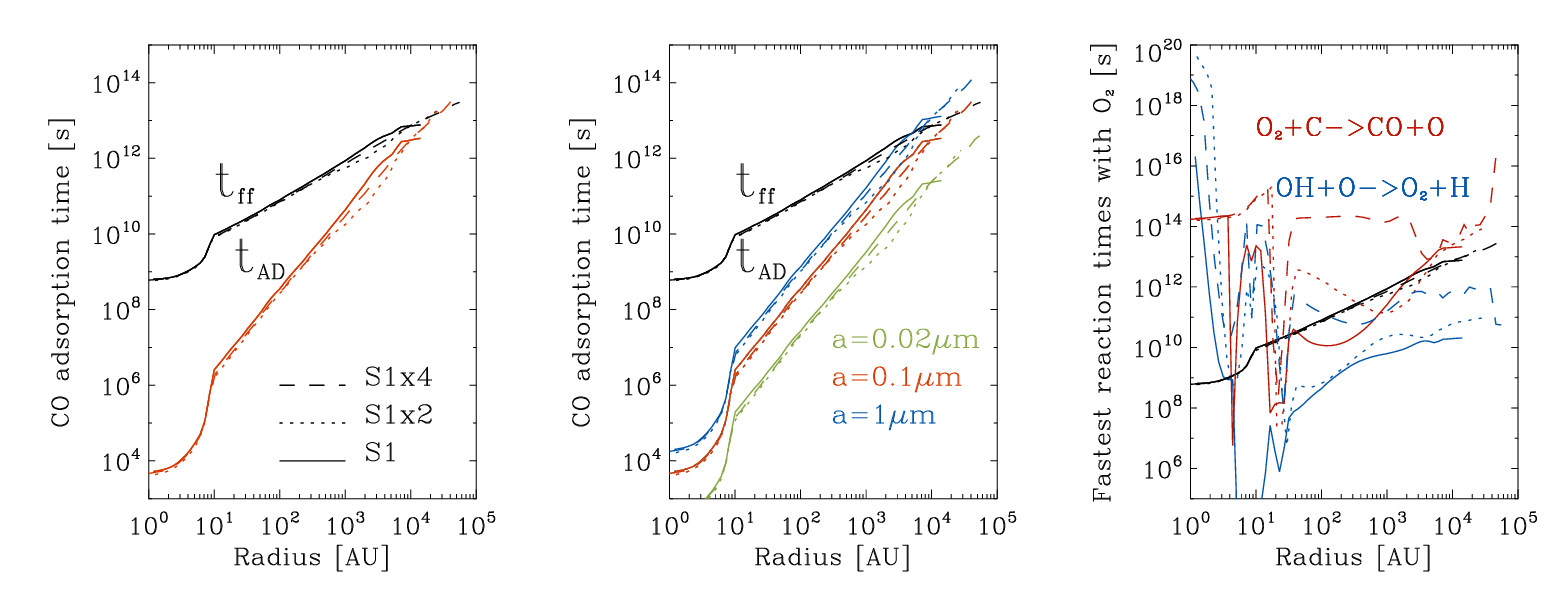}
  \caption{ Comparison between the free-fall time and adsorption time (on example of CO) at the moment of FHSC formation 
        for different free.
            Left: Solid line shows fiducial model (S1), dotted line the model with twice as large box (S1x2), and the
            dashed line the model S1x4 with a box four times larger. 
            Middle: For same models, we show with colors how the CO adsorption time varies with 
              the dust size, where the dust number density is adopted according to Table~\ref{tab:collap}.
            Right: Fastest reactions of creation and destruction of O$_2$, for the same models. } 
\label{box2}
\end{center}
\end{figure*}

According to \citet{lesaffre05} and Eqs. (15)-(16) therein,
the simple estimation of free-fall time and time of CO adsorption on dust grains ($t_\mathrm{AD}$) allows to conclude that 'static chemisorption' of CO should be sufficient for the collapse simulations. 
In fig.~\ref{box2}, we plot the free-fall time and the CO adsorption time for models S1, S1x2, and S1x4 (see Table~\ref{tab:collap}).
As expected, the freeze-out of CO is happening faster than free-fall time almost everywhere in the cloud -- for our choice of parameters. 
Figure ~\ref{box2}, middle, shows the time of CO adsorption on dust grains ($t_\mathrm{AD}$) for various dust sizes.
If the dust has grown to larger sizes (as hinted by observed coreshine) then a dynamical treatment of CO may become necessary in the envelope (fig.~\ref{box2}, $t_\mathrm{ff}<t_\mathrm{AD}$). 

 The main effect of longer free-fall time is found in the behaviour of $\rm O_2$: 
whereas the ices are melted at the same location in all three models, the gas-phase O$_2$ appears at larger radii ($r>20$~AU, outside of FHSC) in the models with the extended radii, S1x2 and S1x4,  than in the fiducial run. In fiducial S1, the gas-phase abundance of $\rm O_2$ shows a sharp decrease dip around 5-6 AU at $t_{0}$. By comparing with other species, we find that a similar behaviour is exhibited by O$_2^+$. At the same location, the abundance of C makes an anti-correlated bump in fiducial model S1. 
As the only difference is the tempo of the collapse, we conclude that this difference occurs due to dynamical effects. In fig.~\ref{box2} we test this conclusion.
The gas-phase reactions are usually slower than the reactions of adsorption-desorption on dust, as shown in fig.~\ref{box2} (right) for the example of O$_2$. We can see that creation of O$_2$ is faster than free-fall time only in the fiducial model S1, and is slower in the extended-box models S1x2 and S1x4. This explains also why there is less gO$_2$ in the fiducial model.%

Intriguingly, there was for a long time a discrepancy between chemical models of the ISM
and $\rm O_2$ non-detections by SWAS and ODIN space telescopes.
 The typical value of $X({\rm O_2})$ in the models was $\sim 10^{-6}$, same as in our fiducial run. 
 The solution to mismatch was to
account for ice chemistry and water formation on grains. When it is included, it consumes
elemental oxygen quite early after the start of chemical evolution in a diffuse ISM,
making elemental O unavailable in large amounts to form significantly abundant molecular oxygen
 ($\rm O_2/H_2 > 10^{-8}$). It is interesting, that a relatively fast collapsing cloud show similar effect - a lack of molecular oxygen whithin a certain radius, pointing out to a second possibility explaining the local absence of O$_2$.


%

Last, the ionization fraction in fig.~\ref{col2} are almost identical for simulations with different initial core radii, because the ionization equilibrium is reached in all simulations. In the center of the cloud, the dominant role of dust grains leads to fast equilibrium too, in spite of the rapidly changing hydrodynamical conditions.

\subsection{Impact of dust and its properties on the chemical abundances in the FHSC \label{sec:dust} }

 In this section, we compare four models with different dust size properties: models S1,
S2$_{\rm MRN}$, S3$_{\rm MRN}$, and S4 (Table~\ref{tab:collap}). S2$_{\rm MRN}$ and S3$_{\rm MRN}$ models make use
of the MRN distribution for the grain sizes in the chemistry module. 
Model S4  has a fixed grain size (as in fiducial model S1) and serve as a boundary case, where we assume very large dust. 
 Note that all the four models use identical set of chemical reactions, initial conditions and dust-to-gas ratio $f_\mathrm{dg}=0.01$, 
 only the dust size properties are varied. 
We refer the reader to Appendix~\ref{sec:ADust} for a more complete description of the treatment of the dust size in {\ttfamily RAMSES-PDS} code. In the following, we briefly summarize the differences between these four models:
\begin{description}
\bf \item[S1]\rm: Fiducial case has single-sized dust with $a_0=0.1\mu$m, 
dust-to-gas ratio, and dust density $n_\mathrm{dust}=3.09\times10^{-12} n_\mathrm{H_2} $;
\item[\bf S2$_{\rm MRN}$\rm]: We use the classical MRN dust grain size distribution $n_{\rm dust}(a)\sim{} a^{-3.5}$ \citep{mathis77},
  scaled to match the results of coreshine modelling \citep{ste10}. The mean dust radius is
  $\sqrt{<a^2>}=0.05$ $\mu$m, what corresponds to $a_\mathrm{min}=2.6\times{} 10^{-6}$ cm and $a_\mathrm{max}=5\times{}10^{-5}$ cm. 
  Fixing $f_\mathrm{dg}=0.01$, the dust number density is calculated to be
   $n_\mathrm{dust}=3.9\times{}10^{-12}n_\mathrm{H_2} $ 
(see Appendix~\ref{sec:ADust} for details);
\item[\bf S3$_{\rm MRN}$\rm]:  Same as S2$_{\rm MRN}$, but  
 we adopt  $a_{\rm min}=10^{-6}$~cm and  $a_{\rm max}=3\times{}10^{-5}$~cm \citep{flower03}. The mean dust radius is $\sqrt{<a^2>}=0.017\mu$m, 
and $n_{\rm dust}=5.24\times{}10^{-11}n_\mathrm{H_2}$. 
This size range resembles the fit for mixed composition 'large' grains, made of bulk carbonaceous material, 
close to the range considered in \citep[][PAH grains are not considered]{mathis77};
\item[\bf S4\rm]: Same as S1, i.e. single-sized dust, with $a_0=1\mu$m, and $n_{\rm dust}=3.09 \times{}10^{-15}n_\mathrm{H_2}$.
%
%
\end{description}

\begin{figure*}
\begin{center}
\includegraphics[width=7.4in]{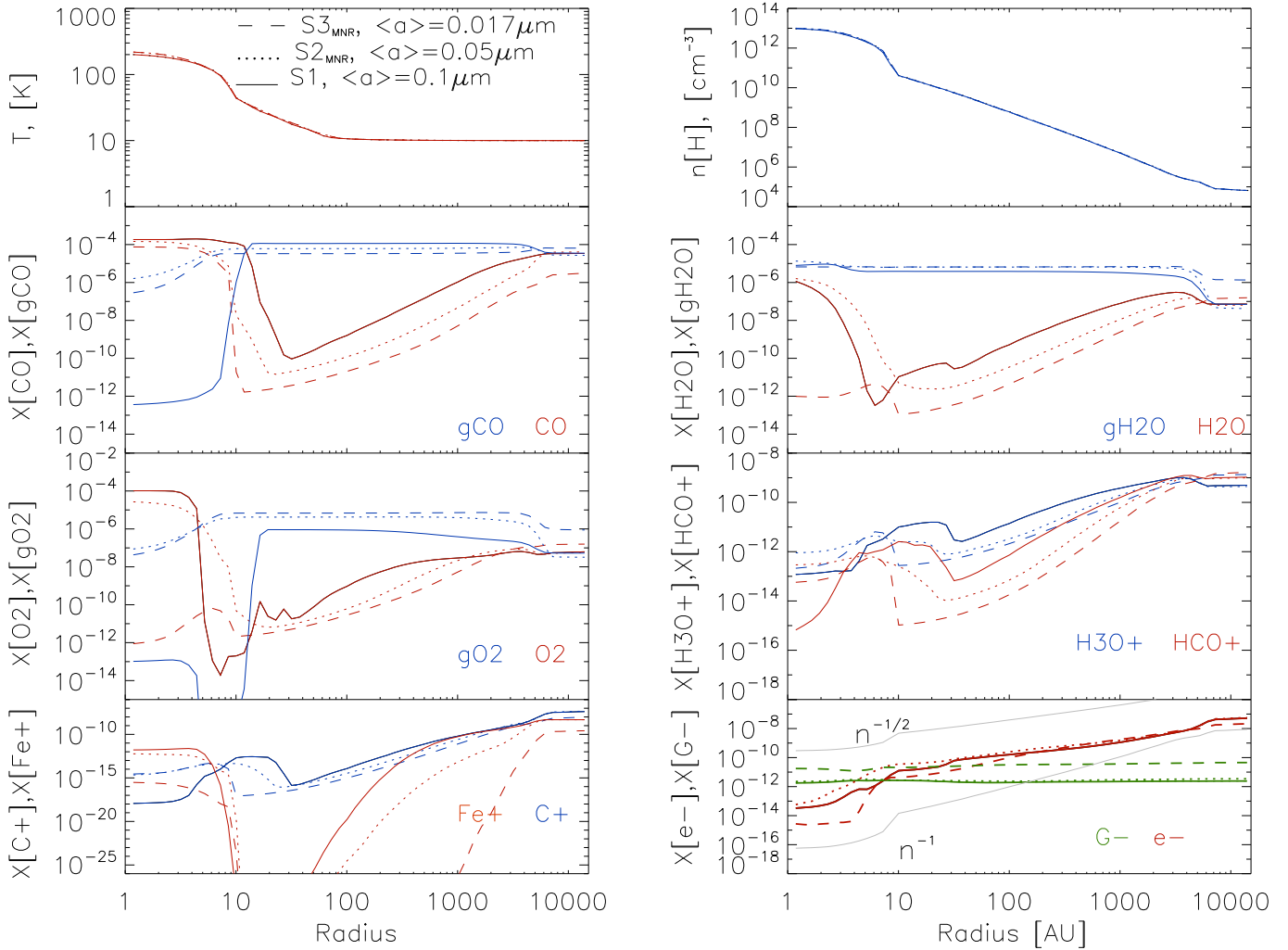}
  \caption{Impact of dust size properties on selected chemical abundances. Same lay-out as in fig.~\ref{col1}. 
            Solid line shows fiducial model (S1), dotted line shows the model S2$_{\rm MRN}$ and dashed line for S3$_{\rm MRN}$, at $t_0$. Abundances from model S5 and S4 are not overplotted for the sake of readability. }
\label{col4}
\end{center}
\end{figure*}

Figure \ref{col4} shows the radial profiles of selected chemical species for models S1, S2$_{\rm MRN}$, and S3$_{\rm MRN}$ at time $t_0$.
The dust properties, such as size distribution and total number density, have a significant impact on the
resulting abundances of all species, especially ions and other charged species.
We observe that the location and steepness of ice sublimation depends on the dust properties too, as visible on the
 plots showing H$_2$O and CO relative abundances: water can already be in the gas phase for fiducial model within the FHSC, but still not melted in S3$_{\rm MRN}$.
The thermal desorption needs indeed more energy to win against faster freeze-out processes in the case of larger dust number density experienced by the  S3$_{\rm MRN}$ model.

\begin{figure}
\begin{center}
\includegraphics[width=3.6in]{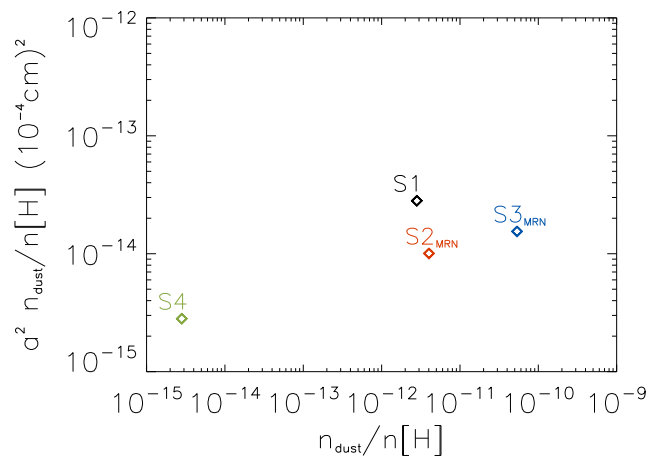}
  \caption{ Mean grain cross-section per hydrogen nucleus as a function od dust density $n_\mathrm{dust}$ for models S1, S4, S5, S2$_{\rm MRN}$ and S3$_{\rm MRN}$. 
Model S2$_{\rm MRN}$ is taken closest to coreshine modelling.
 }
\label{meanopasity}
\end{center}
\end{figure}

In order to understand how dust properties affect the results, we shall concentrate on the reactions of charge transfer 
(eqs.~\ref{chem_solver},~\ref{eq:chtransfer}) and adsorption (eqs.~\ref{chem_solver},~\ref{eq:freezeout}). 
The creation rate in case of charge transfer or recombination is
proportional to $n_\mathrm{dust}$. The creation rate for ice species is proportional to $<a^2>n_\mathrm{dust}$. 
Whereas the dust density decreases steadily from model S1 to S3$_{\rm MRN}$, the parameter  
  $<a^2>n_\mathrm{dust}$ is maximal for S1,
followed by S3$_{\rm MRN}$, with a minimum value in S2$_{\rm MRN}$ (see fig.~\ref{meanopasity}). 
Naively, one could expect that amount of ice species should be maximal in fiducial model S1, as it has largest dust surface available, and an ionization degree should be maximal in  fiducial model S1, followed by S2$_{\rm MRN}$ and S3$_{\rm MRN}$.
We observe that the fiducial model S1 has largest fraction of CO in solid form, 
but in case of water ice and gO$_2$  the 'ice' abundances are maximal for the model S3$_{\rm MRN}$.
Similarly confusing it  is with the amount of electrons: 
We observe that electron fraction can be larger within the FHSC for the case of S2$_{\rm MRN}$, accompanied by larger fraction of $\rm HCO^+$ and $\rm H_3O^+$ compared to fiducial model S1.
Thus the ionization degree, abundances of ions and formation of ices strongly depend on the properties of dust, but to correlate the chemical abundances with the number and with the size distribution of dust is diffucult. 

\begin{figure*}
\begin{center}
\includegraphics[width=7.5in]{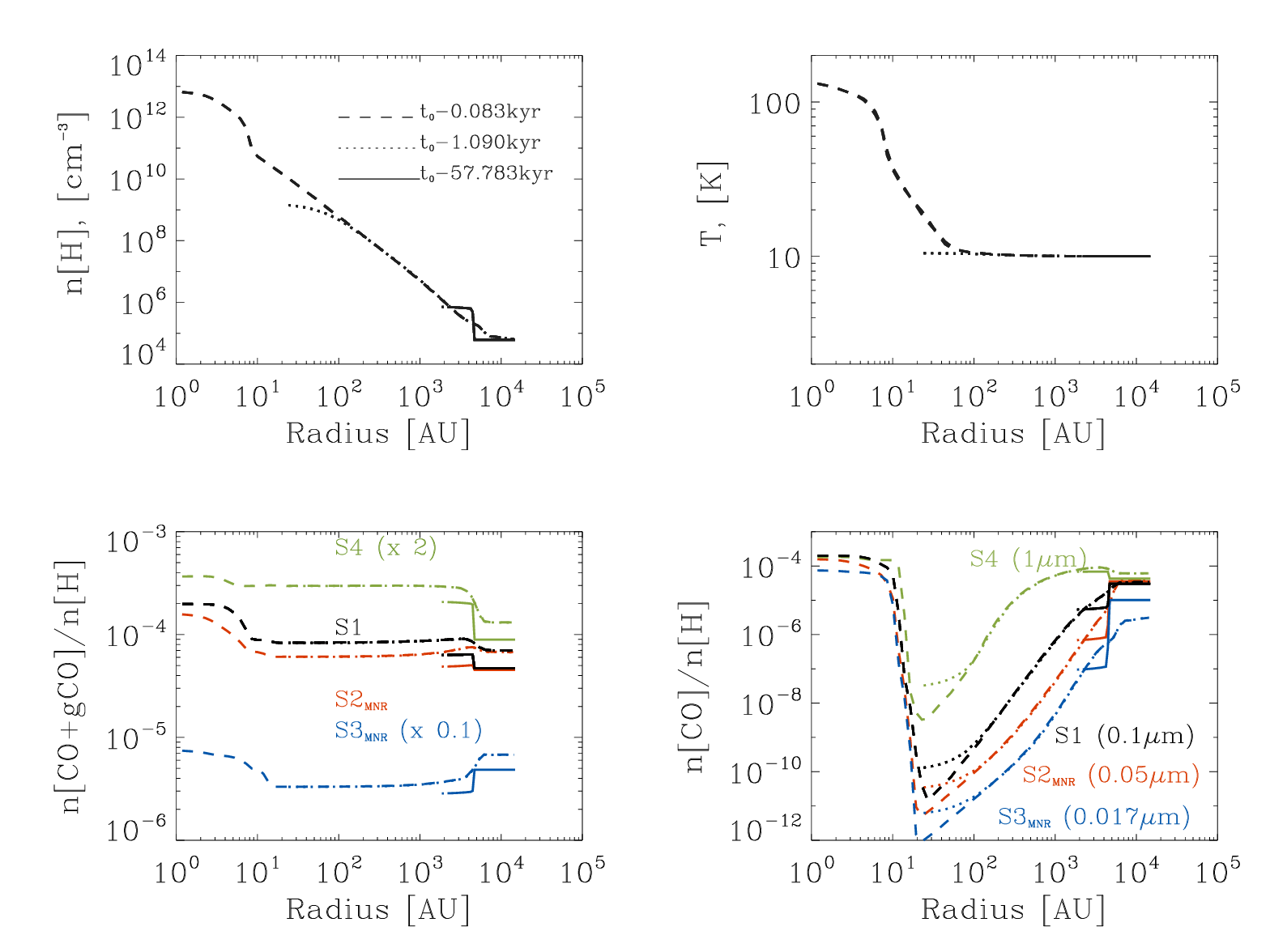}
  \caption{  Radial profiles of gas density, temperature and $X[\rm CO + gCO]$ and $X[\rm CO]$ abundances 
during the collapse starting from initial condition to $t_0$(FHSC formation), 
for  different dust sizes, i.e. models S1, S2$_{\rm MRN}$ 
and S3$_{\rm MRN}$, and S4. 
Note, that the relative total CO abundance is multiplied by a factor 2 in case of S4 and by 0.1 in case of S3$_{\rm MRN}$. 
  }
\label{follow1}
\end{center}
\end{figure*}

Figure \ref{follow1} shows when and how the abundance of CO is affected by the dust size and the dust number density. 
We plot the density and temperature radial profiles for the inital conditions, at time $t_0$ (FHSC formation),
 and at one intermediate time characterizing the isothermal phase of the collapse. 
We show the corresponding radial distribution of gaseous and total CO abundance, $ X[\mathrm{CO}]=(n[\mathrm{CO}]+n[\mathrm{gCO}])/n_\mathrm{H}$ 
 for different dust grain size properties. 
The initial abundances are shown with thick lines. Already at this stage, the radial 
profile of $X[\rm CO]$ differs dramatically with the dust size. In model S2$_{\rm MRN}$, $X[\rm CO]$ depends 
very weakly on the core radius (or gas density), but in models S1 and S4 it increases with gas density,
 and for S3$_{\rm MRN}$ it even decreases with gas density (observe solid lines in fig.~\ref{follow1} bottom left). 

The isothermal phase of the collapse is taking about $5\times10^4$ years in the fiducial case. There is only about thousand years 'left' to reach the FHSC formation during the adiabatic phase of collapse. Thus, the total abundance $X[\rm CO]$ increases during the long (isothermal) phase of collapse due to gas-phase reactions, despite their inefficiency in a dark and poorly ionized cloud. In fig.~\ref{follow1}, $X[\rm CO+gCO]$ is uniform with radius down to the location where the gas is heated up to 25K - 60K. The amount of both gaseous and solid CO differs when the dust properties are changed. 
From this fact and from the comparison displayed in fig.~\ref{follow1} we conclude that the dust properties 
(number, size and distribution slope) have a strong impact on the CO abundance.  

%

We conclude that it is worth to consider the effect of dust growth during the isothermal phase of the collapse
\citep[$T<25$K, $t \leq{}10^{5}$ yrs, e.g.][]{orm09,orm11}.  
After the onset of the adiabatic phase, one can safely work with fixed dust population and concentrate more on 
dynamical chemistry.
Our experiments make clear the importance of the dust modelling for constructing the synthetic collapse CO maps.
The question of dust growth on the timescale of $10^5$ years or longer during the isothermal phase of collapse 
shall be addressed in future work.

\subsection{Limitations of the chemical module}

There are several limitations in our dynamical chemical solver that can potentially affect the estimate of ionisation and thus the resulting magnetic diffusivities. We discuss below the three main limitations and defer their implementation in our model to future work.

First, we assume that the grain size distribution can be represented by a single mean size. \cite{kun09} show that the dust size distribution should be described by at least five bins to have a convergence of less than 1\% in the abundances. 

Second, we consider single-charged grains whereas they can hold multiple electric charges \citep{dra87,dzyur13}. 
Although multiple charged grains are scarce in dense cores \citep{ume80}, they play an important role in the 
ionisation budget.
 In the same line, \cite{mar16} show that charge transfer between grains has a first order effect for densities larger than $10^{10}-10^{11}$ cm$^{-3}$ and that it becomes even more important in the case of multiple charge. Currently, chemical models are limited by numercial difficulties because of the large dynamical range in the abundance of multiple charged grains, which is hard to handle in classical matrix inversion methods such as DVODE. Analytical models have been developed to treat multiple charges but they lack charge transfer between grains  \citep[e.g.][]{dra87}, or force the assumption of the constant ion mass \citep{oku09a}.

Last, we do not account for the variations of the CR ionisation rate deep inside the collapsing core. \cite{Padovani13,pad14} studied the propagation of CR along magnetic fields lines in collapsing cores, accounting for the effect of CR energy loss and magnetic mirroring. 

\section{Conclusion}

We present the first time 3D chemo-dynamical radiation-hydrodynamics simulations of 1 M$_\odot$ isolated dense core collapse. Such calculations are expensive but feasible, as demonstrated in this paper.  The physical
setup includes radiative hydrodynamics and dynamical evolution of a chemical network. In order to perform those
simulations, we merged the multi-dimensional adaptive-mesh-refinement code {\ttfamily{RAMSES}} and 
the thermo-chemistry Paris-Durham shock {\ttfamily (PDS)} code.
We simulate the formation of the first hydro-static core 
 and the co-evolution of 56 species, mainly describing H-C-O chemistry for
the sake of computational feasibility.

The two key aspects of using the chemistry module within the RHD collapse simulation are: First, the observation-oriented aspect of how well the CO abundance can be computed in chemo-dynamical collapse simulations; Second is the theory-oriented aspect of how accurate our computation of magnetic diffusivities can be done in such chemo-dynamical collapse simulations. 
We use a spherical collapse setup and perform a parameter study, to study the dependence of CO abundances on the free-fall time and on the dust size, as well as the effect of dust properties on the ionisation deep inside the core.

In this paper we summarize the findings of the observation-oriented aspect:
\begin{itemize}
\item{} We have systematically tested the reduced chemical network against a well-established full gas-grain network. 
We show that by using a compact set of reactions,
one can get a pretty good match with a much more complex network. This saves computational time and enables the chemo-dynamical RHD simulations of the cloud collapse in 3D. 
\item{}  We follow in detail the time-dependent formation of the first
hydro-static core, until the central temperature of about 800~K and density of about $10^{13}$~cm$^{-3}$
were reached. After that, we use the output physical structure and apply the static gas-grain chemistry 
with an extended set of reactions.
 We find that gas-grain chemistry post-processing can lead to one order
of magnitude lower CO gas-phase abundances compared to the  dynamical chemistry, with strongest effect during the
isothermal phase of collapse. 
\item{}  The duration of the collapse (i.e. free-fall time) has little effect on the chemical abundances for our choice of  parameters.
For mean grain sizes of 1$\mu{}$m and larger, the gas-dust interaction timescales become longer than the representative
dynamical timescales, which affects the pace of molecular depletion and makes dynamical chemistry a must.
\item{} Dust grain mean radius and dust size distribution are  crucial parameters to estimate chemical abundances. Varying the dust size from 0.02 to 0.1 $\mu$m changes the gaseous CO and water abundances by up to 2 orders in magnitude. 
\end{itemize}

In accompanying paper, we concentrate on the effect of the dust properties on the magnetic diffusivities in the cloud. 
Dust mean size and size distribution have a strong effect on chemical
abundances and hence on the ionization degree and magnetic dissipation. 


Dynamical chemical evolution is required to describe  the CO gas phase abundance (as well as the CO ice formation). Linking the chemical evolution to the carefull treatment of the dust properties is of primary importance, regarding the recent observational and theoretical evidences of dust grain growth in the envelope of protostellar cores. Summarizing our finding above, we conclude that proper accounting for
dust grain growth into the collapse simulations can be as important as coupling the collapse
with chemistry.

This proof-of-concept study opens new perspectives for future protostellar collapse numerical models,  in which more complex initial conditions (e.g. rotation and/or turbulence) and self-consistent non-ideal MHD calculations will be investigated. Further chemical process, such as ice formation on the grain surface, should also be considered in the future.


\bibliographystyle{aa}
\bibliography{dzyurkevich1_2016}

\begin{appendix}

\section{Approximate treatment of the dust-size distribution \label{sec:ADust}}

In our chemical module, we want to take into accout the effect of dust-size
distribution on the chemistry and to avoid high memory costs due to treatment of numerous
size bins. 
Here, we trace the number densities only of species G0, $\rm G-$ and $\rm G+$.
For simplicity, we neglect the multiple dust charges.
In case of fixed dust size, the density of dust grain is simply defined as
\begin{equation}
n_\mathrm{dust}=\frac{M_{\rm gas} f_{\rm dg} }{4/3 \pi a^3 \rho_{\rm solid} },
\label{app1}
\end{equation}
where $a$ is the grain radius and $\rho_{\rm solid} $ is the internal density of the solid materials. 
We adopt a MRN size distribution $dn/da \propto a^{-3.5}$ \citep{mathis77}.
Next, we have to choose a range of sizes:
the minimum grain radius is $a_{\rm min}=0.01$ $\mu$m and maximum 
radius $a_{\rm max}=0.3$ $\mu$m. 
For the chemistry, the rates of ion-dust or electron dust reactions are directly affected by 
number density of the dust. The total available
surface of dust grains affects the rates of the freeze-out of species on dust.
We can determine $n_\mathrm{dust}$ using Eq.~\ref{app1}
after calculating the effective cubic radius $R3=<a^3>$ from
\begin{equation}
R3 = \frac{\int_{a_{\rm min}}^{a_{\rm max}}  a^3 a^{-3.5} da}{ \int_{\rm min}^{\rm max} a^{-3.5} da}= 5\frac{a_{\rm max}^{0.5}-a_{\rm min}^{0.5}}
           {a_{\rm min}^{-2.5}-a_{\rm max}^{-2.5}} 
\label{app2}
\end{equation}
Similarly, we can use the number-weighted squared radius $R2=<a^2>$ for reactions sensitive to dust surface, like those in eq.~\ref{eq:freezeout} and eq.~\ref{eq:desorption}. 
\begin{equation}
R2=2.5\frac{a_{\rm min}^{-0.5}-a_{\rm max}^{-0.5}}
            {a_{\rm min}^{-2.5}-a_{\rm max}^{-2.5}}.
\label{app3}
\end{equation}
The average grain radius is 
\begin{equation}
<a>=2.5\frac{a_{\rm min}^{-1.5}-a_{\rm max}^{-1.5}}
            {a_{\rm min}^{-2.5}-a_{\rm max}^{-2.5}}/1.5.
\label{app4}
\end{equation}

Number-weight grain radius and a scale factor between the $<a>$ and $<\sqrt{(R2)}>$ are needed
for correction of the grain cross-section for mantles. Correction factor is 
$a_\mathrm{corr}=<a>/\sqrt{<R2>}$. 

The next input parameter is the average distance between sites on the surface of the grain, 
$\Delta d_\mathrm{sites}=3.4d-8 \ \rm cm$.
Same number is adopted for the thickness of the layers of sites in the grain's mantle.
Number of sites on a single grain of average size is calculated as
\begin{equation}
S=\frac{4\pi <a>^2}{(\Delta d_\mathrm{sites})^2},
\end{equation}
whereas $N_{\rm sites}=(\Delta d_\mathrm{sites})^{-2}$ is a site density in cm$^{-2}$,
taken into account by calculation of the vibrational frequency in eq.~\ref{eq:desorption}.

  \section{Benchmark tests for reduced chemical network
   \label{sec:benching}}

\begin{figure*}
\begin{center}
\includegraphics[width=6.6in]{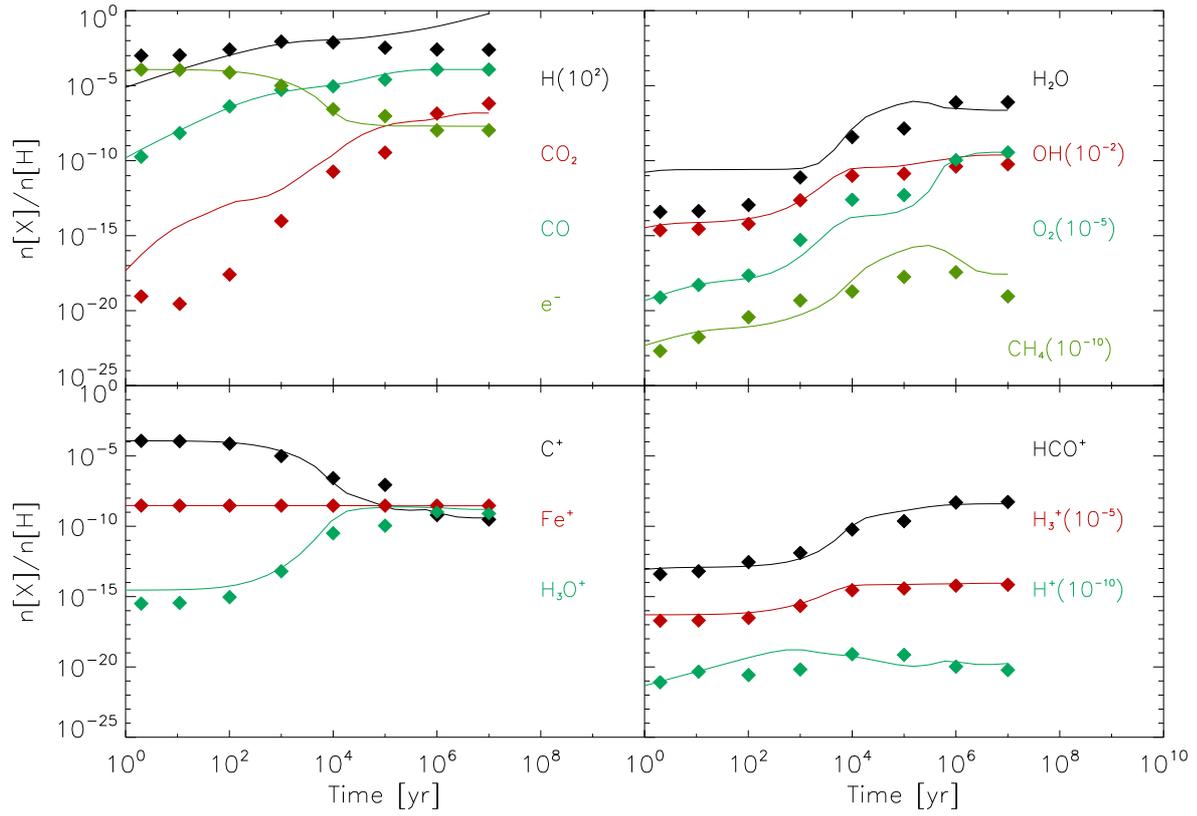}
  \caption{ Time evolution of selected chemical species without dust grain reactions: 
           Solid line shows the case of full chemical network in {\ttfamily ALCHEMIC},
            diamonds the case of reduced network in {\ttfamily RAMSES}.
            }
\label{abumol1}
\end{center}
\end{figure*}

\begin{figure*}
\begin{center}
\includegraphics[width=6.6in]{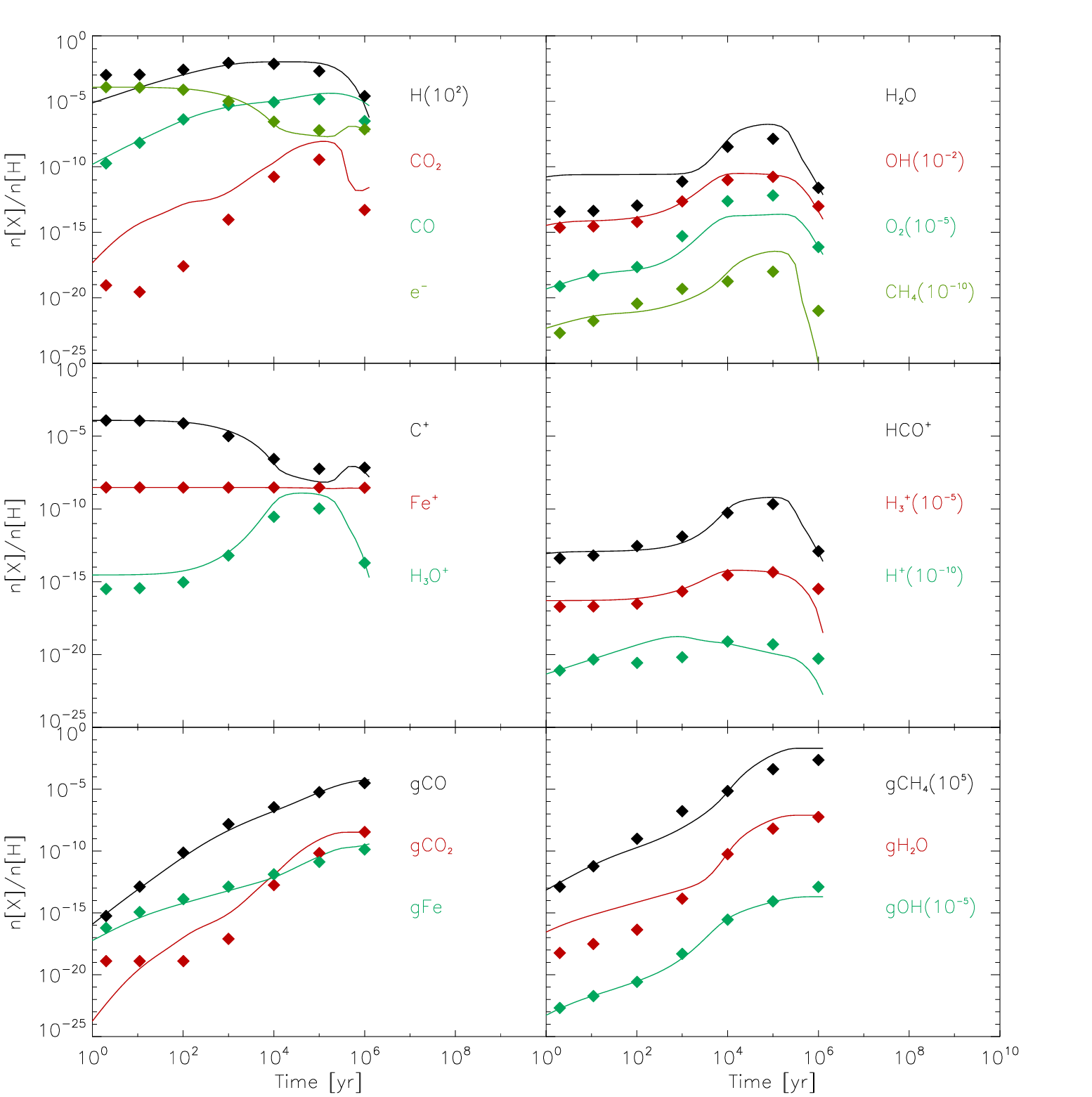}
  \caption{ Time evolution of selected species for gas-phase reactions with freeze-out reactions: 
           Solid line shows full chemical network in {\ttfamily ALCHEMIC}, 
            diamonds the reduced network in {\ttfamily RAMSES}. 
            }
\label{abumol2}
\end{center}
\end{figure*}

\begin{figure*}
\begin{center}
\includegraphics[width=6.6in]{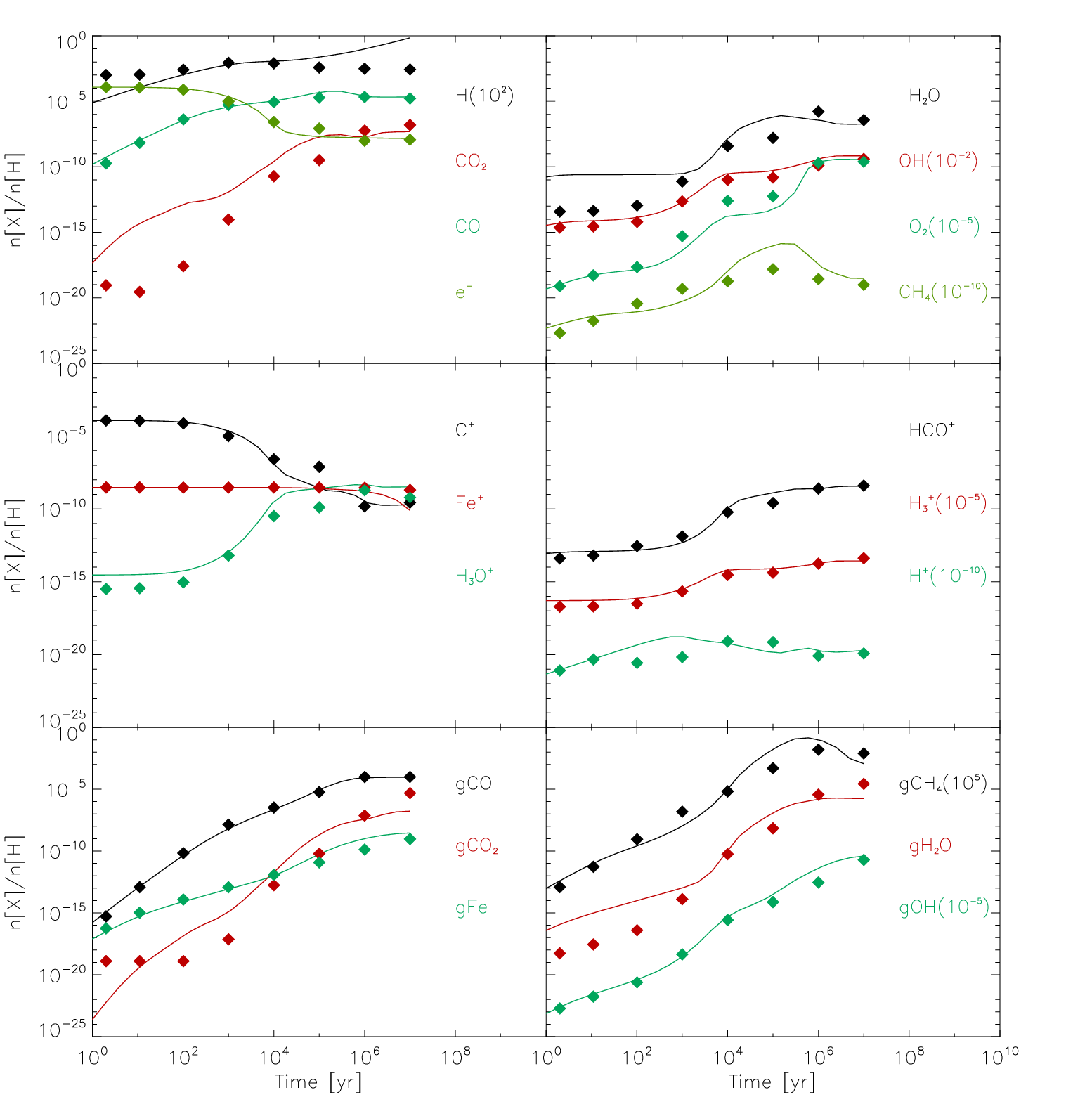} 
  \caption{ Time evolution of selected species for gas-phase chemistry with freeze-out and 
             CR desorption reactions: 
           Solid line shows full chemical network in {\ttfamily ALCHEMIC}, 
            diamonds the reduced network in {\ttfamily RAMSES}. 
            }
\label{abumol3}
\end{center}
\end{figure*}

In this appendix, we compare the results of the full chemical network and of our reduced network in {\ttfamily RAMSES}. First we consider the case of pure gas-chemistry
(fig.~\ref{abumol1}). We then add step-by-step adsorption on the grain surface (fig.~\ref{abumol2}, 
CR-desorption (fig.~\ref{abumol3}). The agreement between both networks is excellent.


\section{Advection test \label{sec:testADV}}

We made minor changes in the \ttfamily{RAMSES}\rm~ code in the advection scheme.  Gas density 
is advected following the continuity equation
\begin{equation}
\frac{\partial \rho}{\partial t} + \nabla(\vec{u}\rho)=0.
\end{equation}
It is possible to define a number of passive scalars in  \ttfamily{RAMSES}\rm, which are always 
considered as relative specie abundances, $n_{specie}/n_{gas}$, and are advected in a different way to the gas density,
\begin{equation}
\frac{\partial \rho}{\partial t} + \vec{u}\nabla(\rho)=0.
\end{equation}
As we want to advect the absolute number density of chemical species, we modified the solver to
treat the $N_\mathrm{species}$ of chemical species in the same way as the gas density, i.e. not as passive scalers.
To test the modifications, we create a simple test. All $N_{\rm species}$ variables for chemistry are replaced with some arbitrary
densities $n=10^{-i+1}n_{\rm gas}$, where $i=0,N_{\rm species}-1$ is a number of variable.
In our test simulation, those densities are advected during the whole length of simulations such as S1 (\ref{tab:collap}).
We have observed that those 'chemical species'
are perfectly advected and keep the inital constant fraction of gas number density during whole collapse simulation 
 to FHSC (first hydro-static core).  

\section{Initial conditions: Abundances of CO for the broad range of gas parameters \label{sec:cube}}

\begin{figure*}
\begin{center}
\includegraphics[width=6.6in]{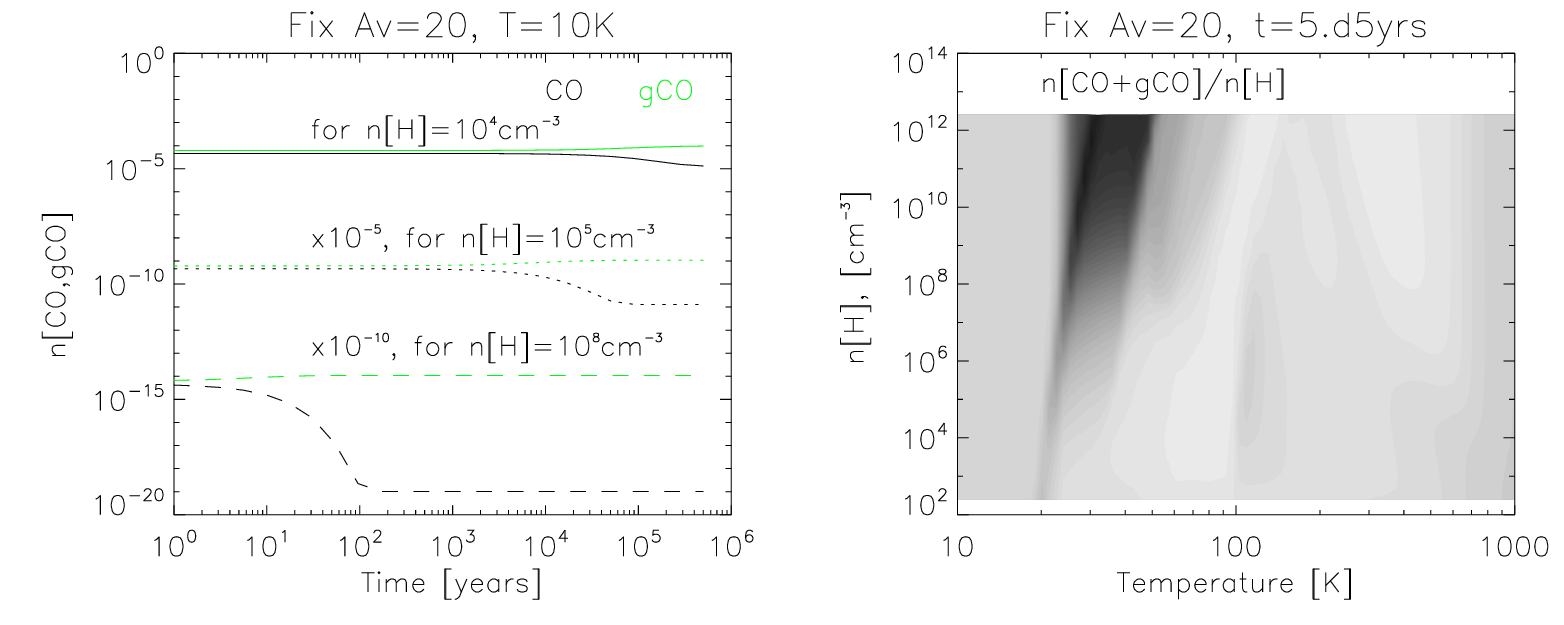} 
  \caption{ Left: Time evolution of CO in gas and solid phases, performed for three different gas densities which are shown in solid, dotted and dashed lines. Right: total CO abundance shown in $(n[H],T)$ space. Visual extinction is fixed to $Av=20$ for both panels. These data are generated with full chemical network {\ttfamily ALCHEMIC} (credits to D. Semenov).
            }
\label{tryCO}
\end{center}
\end{figure*}

Here we present the abundances of CO and its ice, generated by {\ttfamily ALCHEMIC } code for a following set of parameters. For the fixed visual extinktion $Av=20$, the  
temperature ranges from 10K to 1000K, gas number density is avried from $2 \times 10^{2}$ to $2 \times 10^{12} \rm cm^{-3}$. The chemical abundances were evolved up to $5\times{}10^5$yrs for those ranges in density and temperature, which results in the 3D data table. We are interested in how sensitive is the CO abundance to the density of the gas in our initial condition for the collapse. Thus, we plot temporal evolution of relative abundance of CO ($\rm n[CO]/n[H]$, $\rm n[gCO]/n[H]$), for various gas densities and for fixed temperature of $10$K. We observe that the total amount of CO  is independent from the density of the prestellar core, as long as its temperature is not larger than 10K (fig.~\ref{tryCO}). The gaseous CO is less abundant for larger gas densities, and it is faster depleted onto the dust surface for the denser gas (fig.~\ref{tryCO} left, black lines).  

\end{appendix}

\end{document}